\documentclass[useAMS,usenatbib]{mn2e}

\def\aap{AA}

\def\apjl{ApJL}

\def\mnras{MNRAS}
\def\apj{ApJ}
\def\apjs{ApJS}
\def\aj{AJ}

\def\nat{Nat}

\usepackage{bm}
\usepackage{cancel}
\usepackage{graphicx}
\usepackage{float}
\usepackage{amssymb}
\usepackage{amsfonts}
\usepackage{amsmath} 
\usepackage{color}
\usepackage{natbib}
\usepackage{hyperref}
\usepackage{longtable}
\usepackage{caption}

\def\ucla{Department of Physics and Astronomy, PAB, 430 Portola Plaza, Box 951547, Los Angeles, CA 90095-1547, USA}
\def\eso{European Southern Observatory, Karl-Schwarzschild-Strasse 2, 85748 Garching bei M{\"u}nchen, DE}

\def\aaemail{\tt aagnello@eso.org}
\def\pwemail{\tt pwilliams@astro.ucla.edu}

\title[Searching for lensed quasars]{Population mixtures and searches of lensed and extended quasars across photometric surveys} 
\author[Williams, Agnello \& Treu]{
  Peter Williams$^{1,}$\thanks{\pwemail,\aaemail},
  Adriano Agnello$^{1,2},$ and
  Tommaso Treu$^{1}$
  \medskip\\
  $^1$\ucla\\
  $^2$\eso\\
}

\begin{document}

\voffset-.6in

\date{Accepted . Received }

\pagerange{\pageref{firstpage}--\pageref{lastpage}} 

\maketitle

\label{firstpage}

\begin{abstract}
Wide-field photometric surveys enable searches of rare yet interesting objects, such as strongly lensed quasars or quasars with a bright host galaxy. Past searches for lensed quasars based on their optical and near infrared properties have relied on photometric cuts and spectroscopic pre-selection (as in the Sloan Quasar Lens Search), or neural networks applied to photometric samples. These methods rely on cuts in morphology and colours, with the risk of losing many interesting objects due to scatter in their population properties, restrictive training sets, systematic uncertainties in catalog-based magnitudes, and survey-to-survey photometric variations. Here, we explore the performance of a Gaussian Mixture Model to separate point-like quasars, quasars with an extended host, and strongly lensed quasars using $griz$ \texttt{psf} and \texttt{model} magnitudes and WISE $W1,W2.$ The choice of optical magnitudes is due to their presence in all current and upcoming releases of wide-field surveys, whereas UV information is not always available. We then assess the contamination from blue galaxies and the role of additional features such as $W3$ magnitudes or \texttt{psf-model} terms as morphological information. As a demonstration, we conduct a search in a random 10\% of the SDSS footprint, and we provide the catalog of the 43 SDSS object with the highest `lens' score in our selection that survive visual inspection, and are spectroscopically confirmed to host active nuclei. We inspect archival data and find images of 5/43 objects in the Hubble Legacy Archive, including 2 known lenses. The code and materials are available to facilitate follow-up.
\end{abstract}

\section{Introduction}
\label{sect:intro}

Gravitationally lensed quasars offer unique insights into a number of cosmological and astrophysical questions \citep[e.g.,][and references therein]{CSS02}. For example, they can be used to infer properties of the quasar host galaxy \citep{Peng:2006p236}, measure the gravitational profile of lensing object, and probe the nature of dark matter via measurement of substructure. With the addition of time domain information, they can be used for cosmography by using the time-delay between multiple images as a distance indicator \citep[e.g][]{Ref64,Sch++97,K+F99,Ogu07,Suy++10,Suy++13,Suy++14,T+M16}. 

Currently, the main limitation to detailed analyses is the small sample size of known lensed quasars. Only on the order of 100 lenses have been found so far, including only 10-20 of the most valuable kinds like quadruply imaged systems, or highly variable sources. The small sample size is due to the fact that lenses are rare objects and are difficult to find. With the capabilities of current ground based surveys, only about 0.4 are expected per square degree \citep{O+M10}. For example,\footnote{Acronyms of surveys mentioned throughout this paper: 2MASS \citep{Skr06}; DES \citep{DES14}; SDSS \citep{SDSSDR12}; UKIDSS \citep{Hew06}; VHS \citep{VHS13}; VST-ATLAS \citep{VST15}; WISE \citep{WISE10}.}
 in the Dark Energy Survey (DES), 1146 lenses are expected to be found in the 5000 deg$^2$ footprint with roughly 120 of those brighter than 21 in the $i$-band. Of those brighter than 21, about 20\% are expected to be quadruply imaged. Visual inspection in the first 80 deg$^2$ of the Hyper-Suprime Camera \citep[HSC,][]{HSC12} survey has led to the `serendipitous' discovery of a quadruply lensed AGN \citep{Anu16}, owing to a combination of depth and excellent image quality. Similarly, inspection of $\approx6000$ objects in the DES Y1A1 release ($\approx1500\rm{deg}^{2}$), selected as blue point sources near luminous red galaxies, has yielded one new quad lens (Lin et al. 2016, in prep.).

In order to collect a larger lens sample, we need to be able to search through many surveys to pick out potential lens candidates for more detailed follow-up. Due to the size of modern surveys and the lack of readily available spectroscopic data for most objects, we need efficient selection strategies that give a high true positive detection rate based on purely photometric information. Following the strategy and terminology outlined by \citet{Agn++15a}, this selection will result in a smaller and more tractable list of {\it targets}. In turn, those can be subject to computationally more expensive analysis based on the twodimensional survey images to identify {\it candidates} for spectroscopic or high resolution imaging follow-up, with the aim of minimizing time lost to false positives \citep[see also][for an alternative approach]{Sch++16}. This purely photometric strategy differs from the one adopted by the SDSS Quasar Lens Search \citep{Og++06,Ina++12} and its continuation by \citet{Anu16a}, who have relied on objects that already had a confirmed quasar component based on fibre spectra.

The purely photometric selection of targets/candidates can be solved with suitable applications of classification methods in machine learning \citep[see][for a general review of data mining and classification problems in astronomy]{bb10}. One such method is described by \citet{Agn++15a}, where artificial neural networks (ANNs) are used to classify objects as lensed quasars, pairs of closely aligned quasars, alignments of a luminous red galaxy and an unlensed quasar, or blue-cloud galaxies based on their multi-band photometry. This method led to the successful discovery of the first two lensed quasars in DES \citep{Agn++15b}. However, ANNs and similar methods applied to catalog data are very sensitive to systematic differences in the photometry between simulated and real datasets, that can arise from varying image quality conditions or systematics in the creations of catalogs. Furthermore, the machinery cannot be easily carried over from survey to survey, as the photometry of objects (especially those with extended morphology) depends appreciably on survey specifics like image quality, depth, and filter bandpasses. Hence, supervised methods like ANNs require a dedicated training set, which must be tailored closely to the survey being investigated. Another issue is that of extrapolation and generalization, as the properties of objects in different ANN classes reflect directly those that were used as training/validating sets. Finally, ANNs amount to drawing a finite set of hyperplanes in feature space (in our case, colours and magnitudes), governed by the number of nodes chosen, which however cannot be made arbitrarily large before starting to overfit features specific to the training set.

Population mixture models offer several potential advantages with respect to ANNs. Different classes of (known) objects occupy specific regions of colour-magnitude space, so we may model the multi-band properties of objects in wide-field surveys as a superposition of populations, each with its own parent distribution function (PDF). Hence, with a population mixture model one may associate \textit{membership probabilities} to each object in a survey, yielding a smooth (rather than hard-edge) classification. Furthermore, the structural parameters of the PDFs can be fit from the data themselves, allowing for survey-specific adjustments of the classification scheme to account for e.g. different magnitude calibrations. This phenomenon of \textit{augmentation} allows us to initialize the single classes upon a small training set and then adjust them on a large set of un-labelled objects to fit their distribution in feature space.

We will describe the colour-magnitude properties of objects to be classified using a mixture of Gaussian probability distribution functions, whose parameters can be determined recursively with Expectation-Maximization techniques. This approach has been already used in the past for different purposes. It is known alternatively as \textit{Extreme Deconvolution} (XD), as detailed by \citet{bov11a} and applied to quasar identification from point sources in the SDSS. Quasar selection and photometric redshift estimation, among point-sources in the SDSS DR12, has been performed with XD by \citet{DiP15} using optical and WISE magnitudes. A combination of XD and colour cuts was used to select and spectroscopically confirm a sample of $\approx10^4$ quasars in the VST-ATLAS footprint by \citet{che16}. The use of XD tailored to identify lensed quasar targets/candidates has been initiated by Marshall et al.\footnote{https://github.com/drphilmarshall/PS1QLS} for the Panoramic Survey Telescope and Rapid Response System\footnote{Pan-STARRS, http://pan-starrs.ifa.hawaii.edu/public/}. \citet{Fe++16} have discovered at least one new lensed quasar in the DES footprint, among objects classified as quasars through a population mixture approach on optical and infrared magnitudes\footnote{We should point out that, despite the claim of `morphology-independent' data mining, any technique that involves \texttt{psf} and \texttt{model} magnitudes or stellarity is inherently dependent on morphology.}.
In a completely different context, population mixture models have been exploited by \citet{wal09} for the case of multiple stellar populations in nearby dwarf Spheroidal galaxies, then adapted for chemo-kinematic separation of stellar populations and substructure \citep{wp11,kop11,amo12,amo14}, Globular Cluster populations around early-type galaxies \citep{pot13,Agn++14} and substructure detection \citep{lon15}.

This paper is structured as follows. In Section~\ref{sect:comparison} we briefly illustrate the cross-calibration of magnitudes in the SDSS and VST-ATLAS for extended objects. In Section~\ref{sect:popmix} we outline the population-mixture model used to classify the objects in this work. In Section \ref{sect:EMperf}, we examine the performance of our model in SDSS and VST-ATLAS. In Section \ref{sect:candidates}, we present objects selected by the model as potential lens candidates. We conclude in Section~\ref{sect:concl}.

All magnitudes are given in their native system (AB for SDSS and Vega for WISE and VST-ATLAS), and a standard concordance
cosmology with $\Omega_m=0.3$, $\Omega_\Lambda=0.7$, and $h=0.7$ is assumed
when necessary.

\section{Photometry of Extended blue objects in SDSS and VST-ATLAS}
\label{sect:comparison}

Photometry can vary appreciably within one survey and especially from one survey to another due to differences in image quality and depth. This is a worry when methods akin to ANNs are used to classify obects outside of the survey for which they were designed. 

We explore the importance of photometry differences between surveys be examining the cross-calibration of magnitudes in the SDSS and VST-ATLAS, finding the regression that best translates SDSS magnitudes to VST-ATLAS magnitudes. We are interested in both PSF magnitudes and model magnitudes, as one of the key features identifying lensed quasars with image separation comparable or smaller than the seeing is that they have colours similar to quasars but they are extended and therefore should be clearly separable from non-lensed quasars based on the difference between \texttt{model} and \texttt{psf} magnitudes.

A tight fit between SDSS and VST-ATLAS magnitudes is required for ANNs to translate well across the two surveys. Conversely, if there is an appreciable scatter around the regression, this will cause issues for ANNs, illustrating the need for a more robust classification procedure. VST-ATLAS has already been calibrated against SDSS to a precision of $0.02$ magnitudes in zeropoints for point sources, using bright stars \citep[$i<16$,][]{sha15}. In our case, we are interested in the properties of extended and fainter objects, for which the magnitude conversion can differ and scatter can be appreciable.

The SDSS DR12 and VST-ATLAS DR2 (public) footprints overlap mainly in a region with right ascension (r.a.) and declination (dec.) approximately $150^\circ < \text{r.a.} < 230^\circ$ and $-4^\circ < \text{dec.} < -2^\circ$ and one with $-20^\circ < \text{r.a.} < 32^\circ$ and $-12^\circ < \text{dec.} < -10^\circ$. Using matched objects in this region, we can compare the photometry of the two surveys. This will then enable the selection of plausible quasars with extended hosts, or lensed quasar candidates, separating them from classes of contaminants like Seyfert and narrow-line galaxies.

Magnitudes are cross-calibrated between SDSS and VST-ATLAS as follows. We first look at objects that lie within a 5$^{\prime\prime}$ radius of each other in the two surveys. For cross-matched objects, we find that $\Delta\mathrm{r.a.}=-0.08\pm0.81$ and $\Delta\mathrm{dec.}=0.11\pm0.66$ (in arcseconds), where $\Delta\mathrm{r.a.}=\mathrm{r.a.}_{\rm atlas}-\mathrm{r.a.}_{\rm sdss}$ and $\Delta\mathrm{dec.}=\mathrm{dec.}_{\rm atlas}-\mathrm{dec.}_{\rm sdss}$. The distance between matched objects is $\mathrm{dist.} = 0.44 \pm 0.95$ arcseconds. For each band and \texttt{psf} or \texttt{model} magnitude choice, its SDSS and VST-ATLAS counterparts are fit by finding the best offset between the two surveys. In the case of VST-ATLAS, we use the aperture magnitude \texttt{AperMag3} with a 1 arcsecond aperture for our \texttt{psf} magnitudes, and we use \texttt{AperMag6} with a $2\sqrt{2}$ arcsecond aperture for our \texttt{model} magnitudes. The residuals are then fit against adjacent colours, using the maximum likelihood estimator\footnote{Openly available at https://github.com/cristobal-sifon/lnr} \texttt{lnr.mle}, to account for the different shape of the response curves and hence the dependence on the object SED (e.g. quasar or blue galaxy). The resulting regressions for the \texttt{model} magnitudes are

\begin{align}
\nonumber ~&~&~&\text{Intrinsic scatter} \\
\nonumber u_{\rm{atlas}} & = & u_{\rm{sdss}} -0.231(u_{\rm{sdss}}-g_{\rm{sdss}}) - 0.055  &~~~~~~~0.417
\\ 
\nonumber g_{\rm{atlas}} & = & g_{\rm{sdss}} - 0.242 (g_{\rm{sdss}}-r_{\rm{sdss}})+ 0.238  &~~~~~~~0.363
\\ 
\nonumber r_{\rm{atlas}} & = & r_{\rm{sdss}} + 0.042(g_{\rm{sdss}}-r_{\rm{sdss}}) + 0.035   &~~~~~~~0.273
\\ 
\nonumber i_{\rm{atlas}} & = & i_{\rm{sdss}} - 0.005(i_{\rm{sdss}}-z_{\rm{sdss}}) + 0.077  &~~~~~~~0.287
\\ 
 z_{\rm{atlas}} & = & z_{\rm{sdss}}+0.402(i_{\rm{sdss}}-z_{\rm{sdss}}) - 0.084  &~~~~~~~0.284
 \label{eq:modcal}
\end{align}
For the \texttt{psf} magnitudes, we obtain
\begin{align}
\nonumber ~&~&~&\text{Intrinsic scatter} \\
\nonumber g_{\rm{atlas}} & = & g_{\rm{sdss}} - 0.317(g_{\rm{sdss}}-r_{\rm{sdss}})+ 0.158 &~~~~~~~0.343 \\ 
\nonumber r_{\rm{atlas}} & = & r_{\rm{sdss}} -0.577(r_{\rm{sdss}}-i_{\rm{sdss}})+ 0.113 &~~~~~~~0.241 \\ 
i_{\rm{atlas}} & = & i_{\rm{sdss}}-0.108(r_{\rm{sdss}}-i_{\rm{sdss}})-0.021 &~~~~~~~0.247
 \label{eq:psfcal}
\end{align}

The appreciable intrinsic scatter in the translated magnitudes, for the object classes of our interest, is larger than the magnitude uncertainties of single objects.

\section{Population Mixture Models}
\label{sect:popmix}

To deal with issues in translating from SDSS to other surveys or simply from training sets to real data within a survey, we need a technique that is not sensitive to small shifts and scatter in photometry. Population mixture models offer a classification scheme that can be adjusted to the data themselves. In this sense, the model can fine tune itself to fix for any small photometric differences. 

In a population mixture model, we attempt to describe our data as a superposition of parent distribution functions (PDFs), where each PDF captures a different class of objects. For a set of $K$ PDFs each described by parameters $\bm{\theta}_k$, we can construct a log-likelihood function 
\begin{equation}
l(\bm{\theta}) = \log p(\{\bm{x}_i\} | \bm{\theta}) = \log \prod_i^N \sum_k^K p(\bm{x}_i | \bm{\theta}_k),
\end{equation}
where $N$ is the number of objects, $\bm{x}_i$ is a vector of features for each object, and $p(\bm{x}_i | \bm{\theta}_k)$ is the probability that the object $\bm{x}_i$ belongs to class $k$. For computational simplicity, we choose Gaussians as our PDFs, so 
\begin{equation}
p(\bm{x}_i | \bm{\theta}_k) = \frac{\alpha_k}{(2\pi)^{P/2} |\bm{\Sigma}_k |^{1/2}} e^{-\frac{1}{2}(\bm{x}_i - \bm{\mu}_{k})^T\bm{\Sigma}_k^{-1}(\bm{x}_i - \bm{\mu}_{k})}, 
\end{equation}
where $\bm{\mu}_k$ is the mean, $\bm{\Sigma}_k$ is the covariance matrix, $P$ is the number of features, and $\alpha_k$ is a weight parameter defined such that $\sum_k^K \alpha_k = 1$.

In general, one can account for noise in the measurements by convolving with the noise PDF. Since uncertainties are likely dominated by systematic errors such as PSF mismatch and mismatch between the model and the data, the random error is likely a lower limit to the error rather than being representative of the true error. We do not account for noise in this exploration, which is done in the more thorough approach of Extreme Deconvolution \citep{bov11a}. This is appropriate in our context, where random errors are negligible compared to systematic errors associated with model magnitudes, and compared to the intrinsic scatter of the transformations between surveys.

In the case of missing data, we can adjust our model by marginalizing over the missing features. In the case of Gaussian PDFs, this corresponds to restricting $\bm{\Sigma}_k$ and $\bm{\mu}_k$ only to the non-empty entries for each object.

Given this model, our classification problem becomes a matter of finding the parameters $\bm{\theta}$ that maximize the likelihood function. This can be done iteratively using the Expectation Maximization algorithm, briefly described below in the next section and explained in detail in Appendix \ref{sect:EMap}.

The main benefit of using a population mixture model is its flexibility. If two classes are well separated in feature space in our training sets, we expect that they will also be well separated in real survey data, assuming our training sets are good representations of the survey data. Similarly, if they are well separated in one survey, they should be similarly separated in other surveys, despite photometric differences. The changes between different data sets should be only minor changes that can be captured by small adjustments to our parameters, $\bm{\theta}$. Thus, we can train our model in one survey and then let the PDFs adjust themselves via the EM algorithm to translate to other surveys, eliminating the need for separate training sets for each survey.

Further, population mixture models offer a smooth classification of objects in the survey. Each object is assigned a membership probability vector of length $K$, where each entry is the probability that the object belongs to a given class. Finally, adjusting the classes on the whole survey enables a generalization from the objects used in the training (and validating) sets to those that can be met in reality.

\subsection{Expectation-Maximization}
The Expectation Maximization (EM) algorithm is a two-step iterative procedure for finding the parameters that maximize the likelihood function. In the context of our model, we initialize the algorithm with guesses for the parameters $\bm{\theta}$. These can either be random or informed, based on our knowledge of the properties of the classes we wish to describe. In principle, the algorithm should converge to the same parameters regardless of the initial guesses, but as it is often the case in high dimensinoal spaces a good guess helps significantly in guaranteeing rapid convergence to the absolute maximum. In the case of random assignment, we would examine the objects that lie near the mean of each Gaussian at the end of the EM procedure in order to determine which class of objects each Gaussian describes. However, since we know where the objects we wish to categorize lie in feature space, we make this identification at the beginning of the procedure by initializing the parameters based on the expected features for each class.

At each iteration, the \textit{Expectation} step computes the expected value of the log-likelihood function, given the current parameters. This step also computes \textit{membership probabilities}, i.e. a vector for each object giving the probability of belonging to each class. Using these membership probabilities, one can find the parameters that maximize the expected value of the log-likelihood function (i.e. a \textit{Maximization} step). The details on how parameters are updated are described fully in Appendix \ref{sect:EMap}. 

In order to keep parameters from over-adjusting at each step, we introduce a regularization parameter which limits the update of each parameter to a fraction of that proposed by the Maximization step. Since we operate under the assumption that our initial guesses for the parameters are close to the true best parameters, we expect that any updates will be small. By constraining the allowable size of the updates, we avoid any spurious changes to the parameters that might be indicative of one of the Gaussians attempting to encompass multiple classes of objects, or simply due to noise.

In addition, we explored the use of {\it adaptive second moments} to suppress contributions from data lying far from the means of the PDFs. This works by multiplying by an additional windowing factor when calculating the new covariances in the Maximization step. We used a Gaussian centered on the mean of the PDF as the windowing factor so that points farther from the mean would be given less weight. However, this typically led to shrinking the covariances too far to the point that many of the PDF weights went to zero and so we chose not to implement this further.

\subsection{Implementation}

For our particular implementation, we want to distinguish between lensed quasars and various contaminants such as unlensed quasars of different types and blue-cloud galaxies. Typical lensed quasars have the colours of quasars mixed with those of the red lensing galaxy. In particular, their mid-IR colours are similar to those of quasars, while their other colours may be slightly redder. In addition, lenses should be extended and so we expect them to have higher \texttt{psf-model} magnitudes than unlensed QSOs. The value of \texttt{psf-model} will depend on colour and seeing since objects may become deblended in some bands and because the relative contribution of the lens and source vary with wavelength. Blue-cloud galaxies and mergers should have colours similar to QSOs in the optical, but significantly different colours in the mid-IR. 

If we construct features as combinations of colours, then the lenses and contaminants will occupy different locations in features space. Thus, by fitting a PDF to each locus of objects, we can piece together a model describing the overall distribution of our objects. 

\subsubsection{SDSS SpecPhoto training sets}

To develop our training sets, we use objects that are
spectroscopically identified in SDSS SpecPhoto. In particular, we use
objects identified as `QSO' for our unlensed quasar classes and those identified as `GALAXY' for our galaxy class. These will be preferentially brighter than the objects that we wish to classify, but differences in their features should be small. Thus, going from the training sets to real data will require only small adjustments, determined by the EM algorithm. The validity of this hypothesis is quantified in Section~\ref{sect:confusion}.

At different redshifts, quasar spectral features will shift between
filters, intrucing additional complexity to their location in feature
space. To account for this, we break up our quasar classes into six
redshift bins: $z < 0.35$, $0.35 < z < 0.75$, $0.75 < z < 1.2$, $1.2 <
z < 1.75$, $1.75 < z < 2.4$, and $z > 2.4$.

At lower redshifts, we expect some quasar colours to be affected by strong contributions from their host galaxies, making them look more extended. To encapsulate the range of host galaxy contributions, we further break our quasar class into `point-like QSOs' with \texttt{psf - model} magnitudes $<$ 0.15 in $g$, $r$, and $i$ bands, and `extended QSOs' with \texttt{psf - model} $>$ 0.15 in the $i$ band. 

For our galaxy contaminants, we consider only blue-cloud galaxies, selected according to
\begin{align}
\nonumber &u_\text{mod}-r_\text{mod} < 2.2,\\
\nonumber &g_\text{mod}-r_\text{mod} > 0.55 - 0.66(u_\text{mod} - g_\text{mod} - 0.6),\\
\nonumber &g_{\text{psf}} - g_{\text{mod}} > 0.15,\\
\nonumber &r_{\text{psf}} - r_{\text{mod}} > 0.15,\\
&i_{\text{psf}} - i_{\text{mod}} > 0.2
\end{align}
We experimented with including additional galaxy classes, but found that the weights reduced to zero, indicating that they were unnecessary classes that did not describe any new populations in our data.

We use 1000 objects for each class in our training sets. While this will not give the proper weights for the PDFs representing the different classes, it ensures that there are sufficient objects for each class for the machinery to train on. 

\subsubsection{Simulated lensed quasar training sets}
Due to the paucity of known lensed quasars needed to create a full training set, we need to introduce mock lens systems. These are simulated according to the distributions given by \citet{O+M10}. We  further divide the lens class into five separate classes to better capture their diversity.

First, we split the lens objects into those with lower and higher $W1-W2$. Those with lower $W1 - W2$ typically have $i_{\text{psf}} - i_{\text{mod}} < 0.2$ and have colours similar to those of galaxies. These are likely objects with a significant contribution from the lensing galaxy, either due to a larger and brighter galaxy or due to a fainter QSO. 

We break the set of objects with higher $W1-W2$ into four subclasses: 
\begin{itemize}
\item Those with $W1-W2 > 1.5$: These typically have $g - i \sim 1.301$, $i - W1 \sim 3.426$, and $W1 - W2 \sim 1.7$ or higher
\item Those with higher redshift sources: These have $g - i \sim 0.555$, $i - W1 \sim 3.28$, and $W1 - W2 \sim 1.34$
\item Those with lower redshift sources: These have $g - i \sim 0.533$, $i - W1 \sim 4.285$, and $W1 - W2 \sim 1.06$
\item Redder objects: These have $g - i \sim 1.6$, $i - W1 \sim 3.60$, and $W1 - W2 \sim 1.29$
\end{itemize}

In total, our simlated lensed quasar training sets have 2000 objects, 1000 with high $W1-W2$ and 1000 with low $W1-W2$. the simulated lens sample is cut to retain just objects with $W1-W2>0.55,$ where most known lenses lie. 

\begin{figure*}
\centering
\includegraphics[width=6.5in]{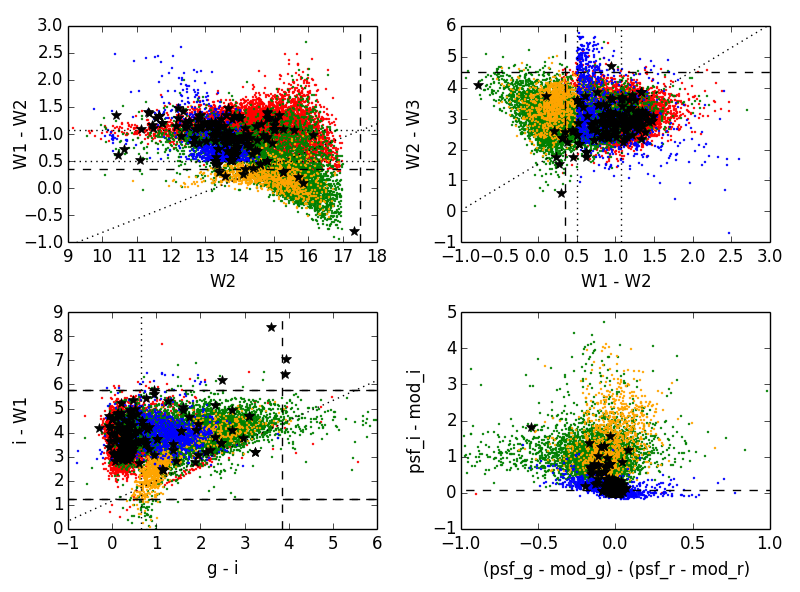}
\caption{Colour-colour plots showing the locations of our training sets in feature space. The red, green, blue, and orange points represent the point-like QSOs, extended QSOs, simulated lensed quasars, and blue cloud galaxies, respectively. The black stars are known lensed quasars that lie in the SDSS footprint. The dashed lines show the colour cuts listed in Equation \ref{eq:colourcuts}. Dotted lines represent either cuts that are within `or' statements or other cuts often used to eliminate contaminants.}
\label{fig:colourcolour}
\end{figure*}

\subsubsection{Large SDSS data set}
\label{sect:largesdss}

After training our model with the training set data, we run the Expectation Maximization algorithm with a large data set. To keep things from becoming too computationally expensive, we first make colour cuts to remove obvious contaminants, and we only take a subset of those objects that satisfy the cuts. The cuts we use are
\begin{eqnarray}
\nonumber &W1 - W2 > 0.35,~W2 - W3 < 4.5,~W2 < 17.5,\\
\nonumber &(W1 - W2 > 0.375 + 0.25\cdot (W2 - 14.76)\\
\nonumber &\text{ or } W2 - W3 < 3.15 + 1.5\cdot(W1 - W2 - 1.075)\\
\nonumber &\text{ or } W1 - W2 > 1.075) \\
\nonumber &1.25 < i_\text{mod} - W1 < 5.75,~i_\text{mod} - W3 < 11.0\\
\nonumber &(g_\text{mod} - i_\text{mod} < 1.2\cdot(i_\text{mod} - W1) - 1.4 \\
\nonumber &\text{ or } g_\text{mod} - i_\text{mod} < 0.65)\\
\nonumber &r_\text{psf} - r_\text{mod} \ge 0.075,~i_\text{psf} - i_\text{model} \ge 0.075 \\
\nonumber &g_\text{mod} - i_\text{mod} < 3.85,~r_\text{mod} - z_\text{mod} < 2.5 \\
&15.0 < i_\text{mod} < 20.5,~ u_\text{mod} - g_\text{mod} < 1.3
\label{eq:colourcuts}
\end{eqnarray}
There are $\sim$6.4 million objects that satisfy these conditions,
which are rather loose. In what follows, we use a random subset of
the top $75\times10^4$ query results. As noted in Table
\ref{tab:colourcuts}, we lose nearly 60 known lenses when we make even
these loose cuts, but these are typically deblended, large separation
lenses. Our search is aimed primarly at the blended small-separation
($\lesssim 2''$)systems, which are the most abundant systems
\citep{Oguri06}, and the least represented in previous searches. 
So these cuts are acceptable for our purposes.

We also see that as we impose even stricter cuts, we can decrease the
number of SDSS objects by an order of magnitude, while keeping nearly
all of the known lenses. We do not use the stricter cuts in this
implementation, but they could be used in the future with larger data
sets to keep things more tractable.

\begin{table}
\centering
\begin{tabular}{l c c}
\hline
Cut	&	Number in SDSS	&	Number of lenses	\\
\hline
Eq. \ref{eq:colourcuts} cuts & $6.4\times 10^6$ & 53 \\
All magnitudes exist & $5.1\times 10^6$ & 53 \\
$i < 20.5,$ $W2 < 15.6$ & $3.6\times 10^6$ & 53 \\
$W1 - W2 > 0.55$ & $2.5\times 10^6$ & 53 \\
Additional (see caption) & $0.5 \times 10^6$ & 50\\
\hline
\end{tabular}
\caption{The number of SDSS objects and known lenses (out of 128) that survive colour cuts increasing in strictness. The SDSS numbers are found by doing a `select count(*)' query, which may count duplicate objects. For this reason, the SDSS values should be treated as order-of-magnitude estimates. The additional colour cuts in the last row are $(u - g < 0.4$ or $g - r < 0.6 - 0.8\cdot (u - g - 0.6)$ or $g - r < 0.4)$. We also note that when we do not include any cuts in \texttt{psf} magnitudes, the number of remaining known lenses increases to 97. When we remove the cuts in $u  - g$ as well, the sample increases to 114.}
\label{tab:colourcuts}
\end{table}

\subsubsection{Choosing features}
\label{sect:feats}

We build our features from combinations of magnitudes in different bands. Figure \ref{fig:colourcolour} shows the distribution of our training set data in varius colour and magnitude spaces. As can be seen, different kinds of objects occupy specific locations in feature space. By choosing features that best distinguish between types of objects and combining them into one high dimensional space, our model can more easily separate the different populations of objects.

The black stars represent 128 known lenses that lie in SDSS, found by various selection techniques. Many have similar colours to point-like quasars, indicative of the colour selection step described by \citet{Og++06}. Others have redder colours, corresponding to either Type-II AGN sources or bright lensing galaxies. We are mainly concerned with finding blended, small-separation lenses, since the large separation counterparts are more likely to have been found already. Therefore, we are less interested in the objects that fall on the point-like QSO locus. 

The dashed lines in the plot show the typical cuts that we use when selecting objects. In order to include lenses that have larger contributions from the lensing galaxy, we use relatively loose cuts in $u-g$ and $W1-W2$. However, known lenses do not lie predominantly at low $W1-W2$ or high $u-g$. Additional metrics, such as those using the $K$ band, can be useful for trimming contaminants in these regimes, as is done by \citet{Fe++16}, but this introduces two issues for our purposes. First, many of the simulated lenses are particularly bright in $K$. This is because the lenses are constructed via sparse interpolation over 2MASS magnitudes, which suffer from Malmquist bias. Second, we are introduced with a depth-vs-footprint trade-off that must be addressed. Surveys such as the Two Micron All-Sky Survey (2MASS) cover the whole sky, but are not always deep enough for our purposes. In fact, only 16\% of our SDSS training set objects have valid $K$ magnitudes in 2MASS. Deeper surveys such as the UKIRT Infrared Deep Sky Survey (UKIDSS) and the VISTA Hemisphere Survey (VHS), on the other hand, cover only a fraction of the SDSS footprint near the equator. 

Finally, we see that most of the known lenses have $i_\text{psf} - i_\text{mod} < 0.8$, with only 10 being more extended. The more extended objects are nearly all lenses with a strong component from a bright lens galaxy. Of the known lenses that do not have $i_\text{psf} - i_\text{mod} \approx 0$, most tend to have $(g_\text{psf} - r_\text{psf}) < (g_\text{mod} - r_\text{mod}),$ indicating that they have a redder extended component. This can be inerpreted as the mixed contribution of the blue (point-like) quasar images and the red (extended) lens galaxy.

For our Gaussian mixture model implementation, we choose to use magnitudes in the SDSS $griz$ bands, avoiding the $u$ band since it is not available in many surveys, for instance, the Dark Energy Survey. In addition, we use the WISE infrared $W1$, $W2$, and $W3$ bands, but exclude the $K$ band for the reasons discussed above. In our `bare-bones' Gaussian mixture model, we use six features: $\texttt{W2}$, $\texttt{mod\_g} - \texttt{mod\_r}$, $\texttt{mod\_g} - \texttt{mod\_i}$, $\texttt{mod\_r} - \texttt{mod\_z}$, $\texttt{mod\_i} - \texttt{W1}$, $\texttt{W1} - \texttt{W2}$. We also experiment with two extensions of the model to include \texttt{W3} magnitudes and \texttt{psf} magnitudes, when available. With the \texttt{W3} magnitudes, we intruduce a $\texttt{W2} - \texttt{W3}$ feature, bringing us to 7 features. The second extension of our model includes $\texttt{psf} - \texttt{model}$ magnitudes as a measure of extendedness. Specifically, we use $\texttt{psf\_i} - \texttt{mod\_i}$, $(\texttt{psf\_g} - \texttt{mod\_g}) - (\texttt{psf\_r} - \texttt{mod\_r})$, and $(\texttt{psf\_r} - \texttt{mod\_r}) - (\texttt{psf\_i} - \texttt{mod\_i})$, bringing us to 9 features.

\subsubsection{Running EM}
\label{sect:running}

We run the EM algorithm in three consecutive steps: First, we take only our lens classes and our lens training sets. Using initial guesses for the parameters that best describe the five classes, we initialize the EM procedure. This then outputs the parameters that maximize the likelihood function. Next, we use all 18 of our classes and include the remainder of our training sets. We use the parameters found in the previous step as our new initial guesses for the lens classes and use our best guesses to initialize the  parameters for the remaining classes. Finally, using the parameters found in the second step, we apply the EM algorithm to the large data sets of different surveys. 

\section{Gaussian Mixture Model Performance}
\label{sect:EMperf}

When training our model, we withhold 30 percent of the objects from the training sets and save them in \textit{validating sets}. We then use these to examine how the model classifies objects that come from the same population as the training sets, but were hidden from the training process. This serves as a check against overtraining, i.e., fitting too closely to the specifics of the training data while failing to accurately represent the class populations as a whole. 

After calculating the membership probabilities of objects in the validating sets, we evaluate the performance of our model by means of confusion matrices and receiver operating characteristic (ROC) curves. Section \ref{sect:confusion} presents the confusion matrices, square matrices showing how objects are classified based on the classes to which they truly belong. These give us insight into how different types of objects are misclassified. Section \ref{sect:roc} shows ROC curves which illustrate how the true positive and false positive lens detection rates change as we vary the acceptance threshold of being a lens, i.e., a minimum lens probability that we use to identify an object as a lens.

\subsection{Confusion Matrices}
\label{sect:confusion}

The confusion matrices are constructed as follows: First, we compute the membership probabilities for all the objects in the validating set, given our output parameters. Next, for each object, we add these probabilities to the cells along the row of the class from which the object truly derives. Finally, we normalize the rows such that the sum of the cells across each row is 1. This, in essence, gives the mean membership probability vector for each class of objects. For a perfect classification scheme, we expect to see ones on the diagonal and zeros elsewhere.

We calculate the confusion matrices at two stages: after running the EM alogrithm with the training set data and after running the algorithm with the full SDSS data set. This allows us to see how the model performs with the real training set data, and then how the performance changes after being `mixed up' by real data. In the latter stage, we look both at what happens when you let only the PDF weights evolve, and then also when you let all parameters evolve. The former is akin to adjusting for the relative abundances of the different class populations, but assuming that the training sets are otherwise perfect representations of the real data. By comparing this to the results of adjusting all parameters, we can determine whether there is anything to be gained from adjusting the means and covariances as well.

\begin{figure*}
\centering
\includegraphics[height=7.5in]{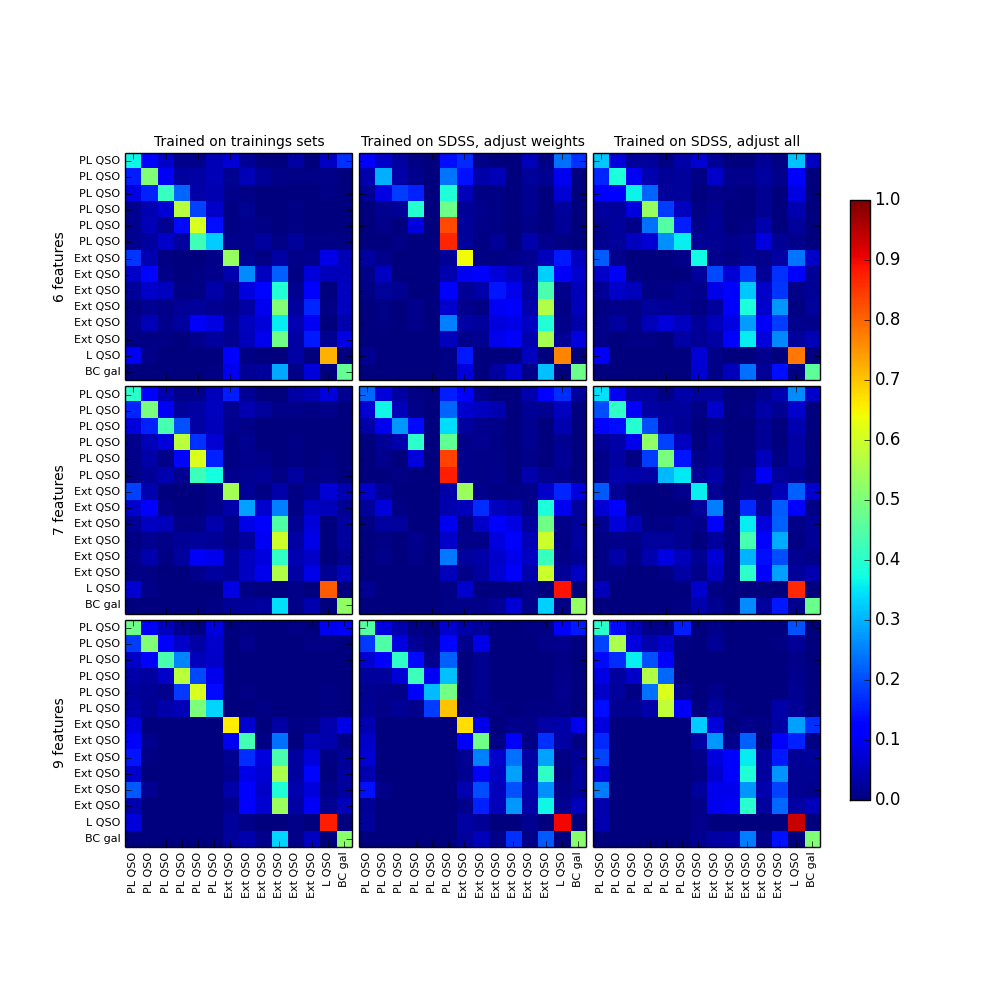}
\caption{Confusion matrices showing how validiating set objects are classified, based on the parameters derived from running the Expectation Maximization algorithm on the training set data. The $y$-axis shows the true class of the input object and the $x$-axis shows to which class the object was identified. The point-like (PL) QSO and extended (Ext) QSO labels are in order of increasing redshift bins from top to bottom on the $y$-axis and left to right on the $x$-axis. The lensed quasar classes are labelled L QSO, and the blue cloud galaxy class is labelled BC gal.}
\label{fig:confusiontrain}
\end{figure*}

The three left panels of Figure \ref{fig:confusiontrain} show confusion matrices generated from the parameters after training on the training set. The top, middle, and bottom panels give the results from the 6, 7, and 9 feature implementations, respectively. We see that the algorithm has difficulties distinguishing between the different `Extended QSO' classes. However, the Point-like QSOs, lensed QSO, and Blue Cloud galaxy classes are well classified. As we add more features, we see further improvement in lens classification. Further, we can note that adding the \texttt{psf\_mag - model\_mag} features helps significantly in distinguishing between the point-like and extended QSOs, as we expected.

The middle column of panels shows the results from running the EM algorithm with the large SDSS data set, but only adjusting the weights, while the rightmost column shows the results after adjusting all parameters. In all three implementations, we see drastic improvement in the classification of point-like QSOs as well as a slight improvement in the classification of all other objects. The improvement demonstrates that adjusting the means and covariances does, indeed, improve the classification abilities of the model beyond that obtained by merely carrying over the same parameters from our training sets to the real data. This step highlights the power of our Gaussian mixture model over more rigid classification schemes.

\subsection{ROC Curves}
\label{sect:roc}

\begin{figure}
\centering
\includegraphics[width=0.3\textheight]{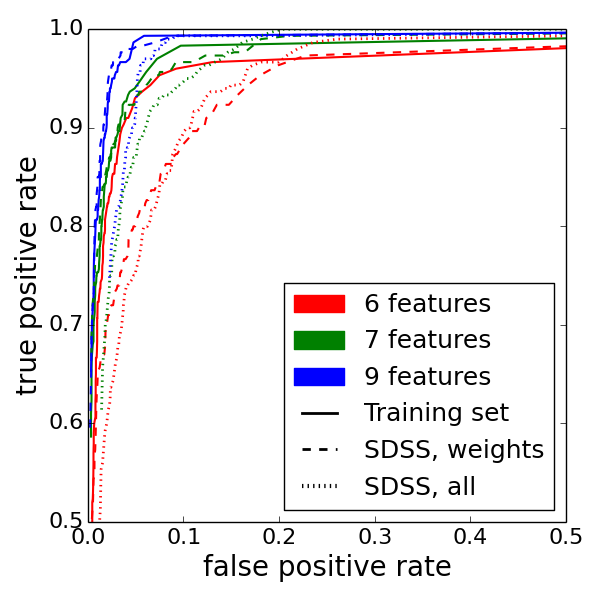}
\caption{Receiver operating characteristic curve, showing the performance of the EM algorithm with the training data. The red, green, and blue lines show the performance based on the 6, 7, and 9 features implementations, respectively. The solid lines show the results from using the training sets, the dashed lines from adjusting only the weights with the SDSD data, and the dotted lines from adjusting all parameters with the SDSS data.}
\label{fig:roccurve}
\end{figure}

Since our end goal is identifying lensed quasars in large surveys, we are interested in the relationship between the true positive and false positive selection rates of our model. Figure~\ref{fig:roccurve} gives the relationship between these two values in the form of a receiver operating characteristic curve. The true and false positive rates are computed by varying the acceptance threshold for identifying a lens. Objects with a combined lens probability above this threshold are identified as lenses and those below the threshold as non-lenses. Given a certain threshold, the true and false positive rates are
\begin{equation}
\text{TPR} = \frac{\text{\# lensing objects identified as lenses}}{\text{\# lensing objects}}
\end{equation}
and 
\begin{equation}
\text{FPR} = \frac{\text{\# non-lensing objects identified as lenses}}{\text{\# non-lensing objects}}.
\end{equation}

If we were to randomly classify each object as a `lens' or 	`nonlens', we would expect TPR and FPR to be identical, corresponding to a line of slope 1 passing through the origin in a ROC curve. The better the performance of the classification scheme, the higher and further left its curve should appear on the diagram. Figure \ref{fig:roccurve} indicates better lens selection rates as we move to models with more features. Note that this shows a zoomed-in portion of the ROC curve with the $x$-axis spanning from 0 to 0.5 and the $y$-axis from 0.5 to 1.0.

\begin{figure}
\centering
\includegraphics[width=3.2in]{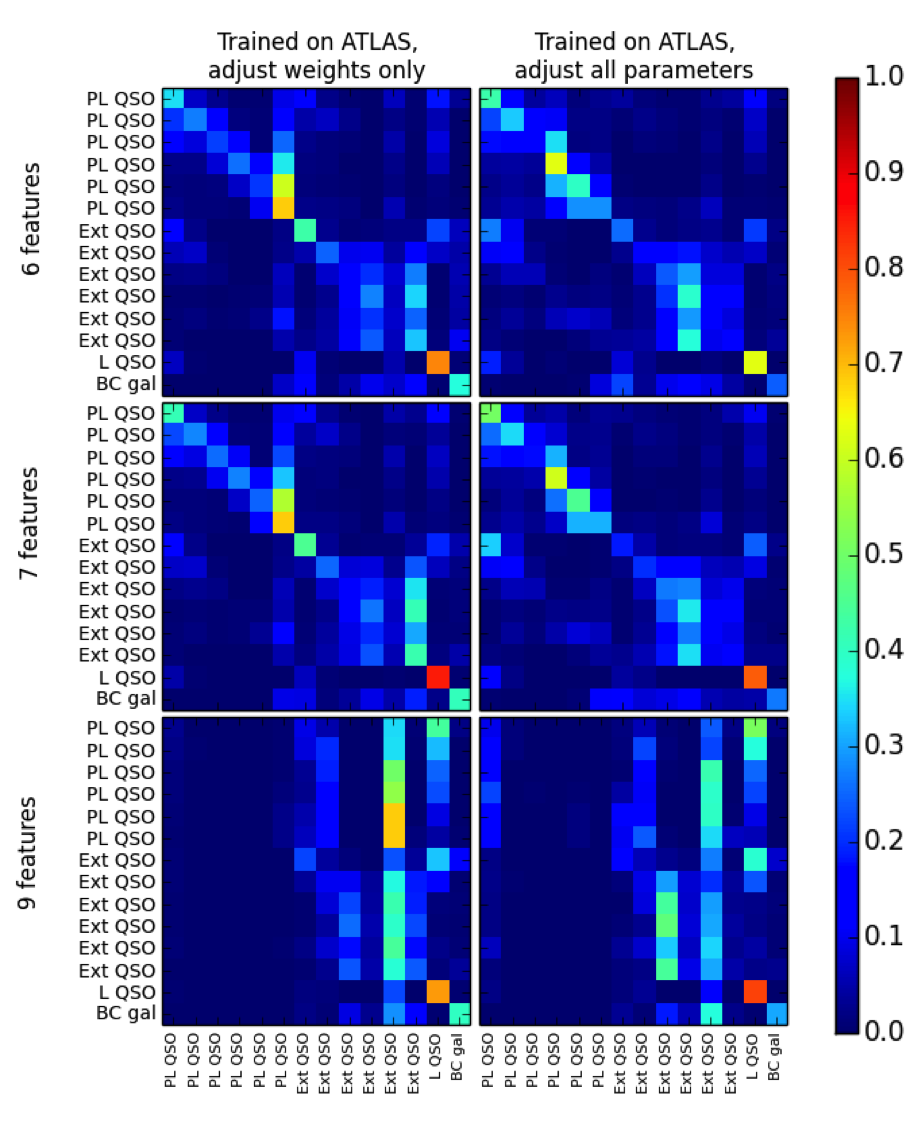}
\caption{Confusion matrices showing the performance of the model when trained on ATLAS data. The point-like (PL) QSO and extended (Ext) QSO labels are in order of increasing redshift bins from top to bottom on the $y$-axis and left to right on the $x$-axis. The lensed quasar classes are labelled L QSO, and the blue cloud galaxy class is labelled BC gal.}
\label{fig:ATLASconfusion}
\end{figure}

\subsection{Testing on SDSS/VST-ATLAS Overlap}

We expect that the PDFs that describe the data from SDSS should be similar to the PDFs that describe the data from other surveys. Only small tweaks should need to be made to the parameters, which can be done using the Expectation Maximization algorithm. We examine this by translating our model trained on the SDSS data to describe similar data in VST-ATLAS.

First, we take the parameters found by training the model on the SDSS data and convert them to VST-ATLAS colours, based on the conversions found in Section \ref{sect:comparison}. If 
\begin{equation}
\mu_\text{atlas} = \bf{A}\mu_\text{sdss}+\bf{B},
\end{equation} where ${\bf A}$ and ${\bf B}$ come from Equations \ref{eq:modcal} and \ref{eq:psfcal}, then
\begin{equation}
\Sigma_\text{atlas} = {\bf A}\Sigma_\text{sdss}{\bf A}^T
\end{equation}
and 
\begin{equation}
\alpha_\text{atlas} = \alpha_\text{sdss}.
\end{equation}
We then use the converted parameters to initialize the EM algorithm with the VST-ATLAS data. After letting the algorithm run, we obtain a new set of parameters to describe the objects in VST-ATLAS.

Since we do not have validating sets in VST-ATLAS, we use the SDSS training sets to evaluate the performance, first converting the SDSS colours to VST-ATLAS colours, again using Equations \ref{eq:modcal} and \ref{eq:psfcal}. The resulting confusion matrices are shown in Figure \ref{fig:ATLASconfusion}. The left column of matrices shows the results when we adjust only the weights. This would be the case if we did not account for differences in photometry and instead fit only for the relative abundance of the populations. The right column shows the results when we adjust all parameters.

We can see that the 6 feature and 7 feature models do well at classifying point-like QSOs, but the performance of the model suffers when the psf-model features are added, which is the opposite of what we see in SDSS. This is likely a sign that unfortunately the VST-ATLAS magnitudes AperMag6 and AperMag3 are not a good proxy for the SDSS model and psf magnitudes.

\subsection{Classification of known lenses in SDSS}
\begin{figure}
\centering
\includegraphics[width=3in]{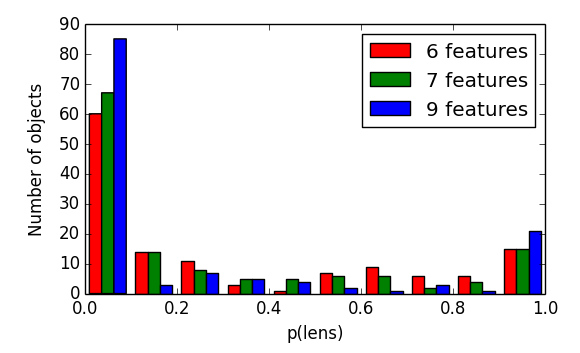}
\caption{Membership probabilities for the 128 known lenses in the SDSS footprint for the 6, 7, and 9 feature models. Note the strong peak at very low and very high probabilities, especially in the 9 feature model.}
\label{fig:knownlensscores}
\end{figure}

We finalize our tests of the Gaussian mixture model by examining how it classifies known lenses. Of a list of known lenses, 128 lie in the SDSS footprint for which we can obtain colours and calculate membership probabilities. Figure \ref{fig:knownlensscores} shows the typical scores assigned to such lenses by the 6, 7, and 9 feature models. The distribution is strongly peaked at the very high probability end and the very low probability end. This is especially pronounced in the 9 feature implementation, where 80 percent of all objects have either $p(\text{lens}) > 0.9$ or $p(\text{lens}) < 0.1$. This shows that the 9 feature implementation is very decisive in making `lens' or `not a lens' assignments. We expect this to be the case as we increase the number of features in the model.

\newcommand{\scalepic}{0.6}
\newcommand{\horizspacing}{0.1in}
\begin{figure}
\begin{flushleft} (a)\end{flushleft}
\centering
\includegraphics[scale=\scalepic]{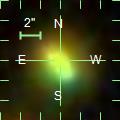}\hspace*{\horizspacing}
\includegraphics[scale=\scalepic]{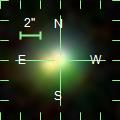}
\hspace*{\horizspacing}
\includegraphics[scale=\scalepic]{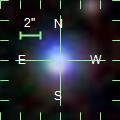}\\
p(lens)=1.0000\hspace*{\horizspacing}p(lens)=0.9915\hspace*{\horizspacing}
p(lens) = 0.9600\\
\vspace*{\horizspacing}
\includegraphics[scale=\scalepic]{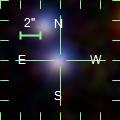}\hspace*{\horizspacing}
\includegraphics[scale=\scalepic]{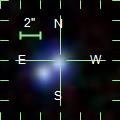}
\hspace*{\horizspacing}
\includegraphics[scale=\scalepic]{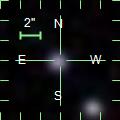}\\
p(lens) = 0.0501\hspace*{\horizspacing}
p(lens) = 0.0135\hspace*{\horizspacing}
p(lens) = 0.0002\\
\vspace*{0.1in}
\begin{flushleft} (b)\end{flushleft}
\vspace*{-12pt}
\includegraphics[width=3in]{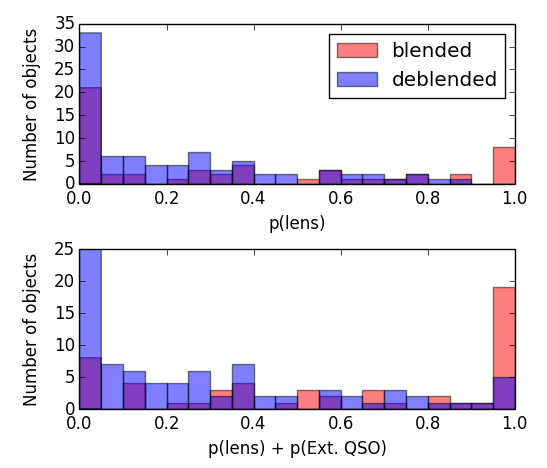}
\caption{(a): Image cutouts of known lensed quasars with the lens probabilities assigned by our model. (b): Membership probabilities for the 128 known lenses, divided into `blended' and `deblended' categories. `Blended' objects look like those in the the top three images of part (a) while `deblended' objects look like those in the bottom three images of part (a). The $x$-axis gives the lens probability and the combined lens and extended QSO probabilities in the top and bottom plots, respectively.}
\label{fig:knowncutouts}
\end{figure}

Examining the SDSS image cutouts, we can see a pattern that the objects with the highest lens probability are typically blended, while those with the lowest probabilities are often very well separated. A select few cutouts are shown in Figure \ref{fig:knowncutouts} (a), along with their $p$(lens) score. In Figure \ref{fig:knowncutouts} (b), we show the score distribution for deblended and blended objects. The top plot displays only the lens probability while the bottom plot uses the sum of the lens probability and the extended QSO probabilities. Approximately one-third of the blended objects have a combined lens and extended QSO probability greater than 0.95. Similarly, approximately one-third of the deblended objects have a combined lens and extended QSO probability less than 0.05, with the highest scores coming from the point-like QSO classes. The complete list of lenses is given in Table \ref{tab:knownlensscores}, along with selected scores. We note that, owing to selection effects, wide separation systems are over-represented in the list of known lenses, with respect to what is expected in nature.

\begin{table*}
\centering
\setlength{\tabcolsep}{4.0pt}
{\renewcommand{\arraystretch}{0.98}
\begin{tabular}{c r r r c c c c c}
\hline																	
Name	&	ra (deg)	&	dec (deg)	&	p(lens)	&	$\begin{array}{c}p\text{(PL QSO)}\\\text{all redshifts}\end{array}$	&	$\begin{array}{c}p\text{(Ext QSO})\\ 1.2 < z < 1.75\end{array}$	&	$\begin{array}{c}p\text{(Ext QSO})\\ 1.75 < z < 2.4\end{array}$	&	$\begin{array}{c}p\text{(Ext QSO})\\ z > 2.4\end{array}$	&	Sep ('')	\\
\hline													
Q0015+0239	&	4.5474477	&	2.9444539	&	0.00014	&	0.99772	&	0.00000	&	0.00206	&	0.00000	&	2.20	\\
TEX0023+171A	&	6.4040475	&	17.4680462	&	0.00278	&	0.48241	&	0.00000	&	0.00779	&	0.00001	&	4.80	\\
SDSSJ00480-1051A	&	12.0032122	&	-10.8634797	&	0.06170	&	0.93825	&	0.00000	&	0.00002	&	0.00000	&	3.60	\\
PMNJ0134-0931	&	23.6486210	&	-9.5175140	&	0.99999	&	0.00001	&	0.00000	&	0.00000	&	0.00000	&	0.73	\\
Q0142-100	&	26.3191555	&	-9.7548180	&	0.69409	&	0.30590	&	0.00000	&	0.00000	&	0.00000	&	2.24	\\
PHL1222	&	28.4745284	&	5.0491993	&	0.46181	&	0.53818	&	0.00000	&	0.00000	&	0.00000	&	3.30	\\
SDSSJ02452-0113A	&	41.3005190	&	-1.2205661	&	0.00071	&	0.99840	&	0.00000	&	0.00087	&	0.00000	&	4.50	\\
SDSS0246-0825	&	41.6420778	&	-8.4267136	&	0.36548	&	0.63451	&	0.00000	&	0.00000	&	0.00000	&	1.19	\\
SDSSJ02483+0009A	&	42.0866741	&	0.1657625	&	0.00038	&	0.99953	&	0.00000	&	0.00005	&	0.00000	&	6.90	\\
MG0414+0534	&	63.6573133	&	5.5784262	&	0.99779	&	0.00000	&	0.00000	&	0.00221	&	0.00000	&	2.40	\\
B0445+123	&	72.0918766	&	12.4654827	&	0.23227	&	0.00000	&	0.00000	&	0.64398	&	0.00229	&	1.35	\\
SDSSJ07402+2926A	&	115.0560383	&	29.4467821	&	0.57241	&	0.42759	&	0.00000	&	0.00000	&	0.00000	&	2.60	\\
SDSS J0743+2457	&	115.9692451	&	24.9621262	&	0.00408	&	0.65394	&	0.00002	&	0.01489	&	0.00093	&	1.03	\\
SDSS0746+4403	&	116.7210121	&	44.0642878	&	0.06333	&	0.88548	&	0.00000	&	0.00077	&	0.00003	&	1.11	\\
SDSSJ07479+4318A	&	116.9959352	&	43.3014976	&	0.01982	&	0.97833	&	0.00000	&	0.00001	&	0.00000	&	9.20	\\
SDSS0806+2006	&	121.5987673	&	20.1088663	&	0.03328	&	0.96555	&	0.00000	&	0.00001	&	0.00000	&	1.40	\\
HS0810+2554	&	123.3803634	&	25.7508530	&	0.65535	&	0.34465	&	0.00000	&	0.00000	&	0.00000	&	0.96	\\
SDSSJ08199+5356	&	124.9992399	&	53.9400507	&	0.65332	&	0.00000	&	0.07066	&	0.00041	&	0.00313	&	4.04	\\
SDSSJ08202+0812	&	125.0671257	&	8.2044367	&	0.00219	&	0.99698	&	0.00000	&	0.00002	&	0.00000	&	2.30	\\
HS0818+1227	&	125.4122634	&	12.2916624	&	0.10073	&	0.89729	&	0.00000	&	0.00092	&	0.00000	&	2.10	\\
APM08279+5255	&	127.9237751	&	52.7548742	&	0.99989	&	0.00011	&	0.00000	&	0.00000	&	0.00000	&	0.38	\\
SDSS J0832+0404	&	128.0708254	&	4.0681127	&	0.37806	&	0.62059	&	0.00000	&	0.00006	&	0.00000	&	1.98	\\
SDSS0903+5028	&	135.8955917	&	50.4720356	&	0.86865	&	0.00000	&	0.00000	&	0.00024	&	0.00027	&	2.99	\\
SDSS J0904+1512	&	136.0173140	&	15.2151511	&	0.37964	&	0.62023	&	0.00000	&	0.00000	&	0.00000	&	1.13	\\
RXJ0911+0551	&	137.8650616	&	5.8483616	&	0.26514	&	0.73404	&	0.00000	&	0.00073	&	0.00000	&	2.47	\\
SBS0909+523	&	138.2542954	&	52.9913607	&	0.99151	&	0.00849	&	0.00000	&	0.00000	&	0.00000	&	1.17	\\
RXJ0921+4529	&	140.3034237	&	45.4844531	&	0.01664	&	0.98311	&	0.00000	&	0.00001	&	0.00000	&	6.97	\\
SDSS0924+0219	&	141.2325339	&	2.3236837	&	0.37140	&	0.62585	&	0.00000	&	0.00001	&	0.00000	&	1.75	\\
SDSS J0946+1835	&	146.5199732	&	18.5943605	&	0.00491	&	0.00000	&	0.17294	&	0.00226	&	0.27944	&	3.06	\\
FBQ0951+2635	&	147.8440691	&	26.5872082	&	0.72388	&	0.27612	&	0.00000	&	0.00000	&	0.00000	&	1.11	\\
BRI0952-0115	&	148.7503824	&	-1.5018700	&	0.00320	&	0.00011	&	0.10729	&	0.03752	&	0.07157	&	1.00	\\
SDSSJ09591+5449A	&	149.7811562	&	54.8184576	&	0.00001	&	0.99991	&	0.00000	&	0.00007	&	0.00000	&	3.90	\\
Q0957+561	&	150.3368322	&	55.8971172	&	0.42382	&	0.57618	&	0.00000	&	0.00000	&	0.00000	&	6.26	\\
SDSS1001+5027	&	150.3692164	&	50.4657960	&	0.55823	&	0.44177	&	0.00000	&	0.00000	&	0.00000	&	2.82	\\
J1004+1229	&	151.1037295	&	12.4895218	&	0.99618	&	0.00002	&	0.00000	&	0.00375	&	0.00000	&	1.54	\\
SDSS1004+4112	&	151.1455147	&	41.2118874	&	0.02993	&	0.96983	&	0.00000	&	0.00000	&	0.00000	&	15.99	\\
SDSS1011+0143	&	152.8729178	&	1.7231710	&	0.00144	&	0.00000	&	0.41954	&	0.00155	&	0.28144	&	3.67	\\
LBQS1009-0252	&	153.0650962	&	-3.1169016	&	0.08552	&	0.91415	&	0.00000	&	0.00024	&	0.00000	&	1.54	\\
SDSS1021+4913	&	155.2959315	&	49.2250908	&	0.33645	&	0.65626	&	0.00000	&	0.00009	&	0.00000	&	1.14	\\
FSC10214+4724	&	156.1440054	&	47.1526938	&	0.05473	&	0.51565	&	0.00000	&	0.00042	&	0.00000	&	1.59	\\
SDSSJ10287+3929A	&	157.1819591	&	39.4935992	&	0.00024	&	0.98231	&	0.00000	&	0.01161	&	0.00002	&	7.50	\\
B1030+074	&	158.3917723	&	7.1905771	&	0.38193	&	0.61359	&	0.00000	&	0.00001	&	0.00000	&	1.65	\\
SDSSJ10353+0752A	&	158.8306849	&	7.8827901	&	0.06756	&	0.93243	&	0.00000	&	0.00001	&	0.00000	&	2.70	\\
SDSS J1054+2733	&	163.6701660	&	27.5517870	&	0.72399	&	0.27601	&	0.00000	&	0.00000	&	0.00000	&	1.27	\\
SDSS J1055+4628	&	163.9393929	&	46.4776390	&	0.13538	&	0.85958	&	0.00000	&	0.00142	&	0.00021	&	1.15	\\
SDSSJ10567-0059A	&	164.1870081	&	-0.9926159	&	0.00000	&	0.99815	&	0.00000	&	0.00171	&	0.00001	&	7.20	\\
HE1104-1805	&	166.6399556	&	-18.3569516	&	0.64122	&	0.35878	&	0.00000	&	0.00000	&	0.00000	&	3.19	\\
SDSSJ11161+4118A	&	169.0488890	&	41.3059807	&	0.37839	&	0.49905	&	0.00000	&	0.04115	&	0.00058	&	13.00	\\
PG1115+080	&	169.5706277	&	7.7661757	&	0.79756	&	0.20244	&	0.00000	&	0.00000	&	0.00000	&	2.32	\\
SDSSJ11202+6711	&	170.0504960	&	67.1877758	&	0.03567	&	0.92827	&	0.00000	&	0.00097	&	0.00001	&	1.50	\\
UM425	&	170.8363735	&	1.6298543	&	0.95997	&	0.04003	&	0.00000	&	0.00000	&	0.00000	&	6.50	\\
SDSSJ11249+5710A	&	171.2302020	&	57.1823829	&	0.32289	&	0.66965	&	0.00000	&	0.00437	&	0.00000	&	2.20	\\
SDSS J1128+2402	&	172.0770482	&	24.0381996	&	0.12290	&	0.87701	&	0.00000	&	0.00000	&	0.00000	&	0.84	\\
SDSS J1131+1915	&	172.9905204	&	19.2576997	&	0.23650	&	0.76314	&	0.00000	&	0.00031	&	0.00000	&	1.46	\\
SDSS1138+0314	&	174.5155742	&	3.2493912	&	0.01296	&	0.95709	&	0.00000	&	0.00040	&	0.00002	&	1.34	\\
SDSSJ11381+6807A	&	174.5383643	&	68.1274026	&	0.18903	&	0.81097	&	0.00000	&	0.00000	&	0.00000	&	2.60	\\
SDSS1155+6346	&	178.8222976	&	63.7727990	&	0.04832	&	0.00050	&	0.00002	&	0.00002	&	0.00002	&	1.95	\\
B1152+200	&	178.8262178	&	19.6617257	&	0.58448	&	0.41552	&	0.00000	&	0.00000	&	0.00000	&	1.59	\\
SDSSJ11583+1235A	&	179.5948971	&	12.5884958	&	0.00432	&	0.98312	&	0.00000	&	0.00003	&	0.00000	&	3.60	\\
SDSS1206+4332	&	181.6235366	&	43.5382141	&	0.12480	&	0.87505	&	0.00000	&	0.00000	&	0.00000	&	2.90	\\
1208+1011	&	182.7376442	&	9.9074865	&	0.60758	&	0.30387	&	0.00063	&	0.07242	&	0.01046	&	0.45	\\
SDSSJ12167+3529	&	184.1918625	&	35.4948606	&	0.00195	&	0.99741	&	0.00000	&	0.00007	&	0.00000	&	1.50	\\
HS1216+5032A	&	184.6708726	&	50.2599576	&	0.53323	&	0.46677	&	0.00000	&	0.00000	&	0.00000	&	8.90	\\
SDSSJ12257+5644A	&	186.4405459	&	56.7446074	&	0.00076	&	0.97687	&	0.00001	&	0.01671	&	0.00004	&	6.00	\\
\hline
\end{tabular}}
\caption{\textit{Continued on next page.}}
\end{table*}
\addtocounter{table}{-1}

\begin{table*}
\centering
\setlength{\tabcolsep}{4.0pt}
{\renewcommand{\arraystretch}{0.98}
\begin{tabular}{c r r r c c c c c}
\hline
Name	&	ra (deg)	&	dec (deg)	&	p(lens)	&	$\begin{array}{c}p\text{(PL QSO)}\\\text{all redshifts}\end{array}$	&	$\begin{array}{c}p\text{(Ext QSO})\\ 1.2 < z < 1.75\end{array}$	&	$\begin{array}{c}p\text{(Ext QSO})\\ 1.75 < z < 2.4\end{array}$	&	$\begin{array}{c}p\text{(Ext QSO})\\ z > 2.4\end{array}$	&	Sep ('')	\\
\hline
SDSS1226-0006	&	186.5334332	&	-0.1006261	&	0.64129	&	0.35862	&	0.00000	&	0.00002	&	0.00000	&	1.26	\\
SDSSJ12511+2935	&	192.7815624	&	29.5945841	&	0.76803	&	0.18325	&	0.00000	&	0.00001	&	0.00000	&	1.79	\\
SDSSJ12543+2235	&	193.5789495	&	22.5934873	&	0.03605	&	0.26113	&	0.00314	&	0.07851	&	0.04694	&	1.56	\\
SDSSJ12583+1657	&	194.5801292	&	16.9549317	&	0.25021	&	0.74965	&	0.00000	&	0.00006	&	0.00000	&	1.28	\\
SDSSJ12599+1241A	&	194.9817306	&	12.6982776	&	0.00006	&	0.99936	&	0.00000	&	0.00023	&	0.00000	&	3.60	\\
SDSSJ13034+5100A	&	195.8590567	&	51.0131162	&	0.00183	&	0.99740	&	0.00000	&	0.00070	&	0.00000	&	3.80	\\
SDSS J1304+2001	&	196.1816022	&	20.0178274	&	0.01713	&	0.98218	&	0.00000	&	0.00002	&	0.00000	&	1.87	\\
SDSSJ13136+5151	&	198.4166011	&	51.8579110	&	0.29308	&	0.70692	&	0.00000	&	0.00000	&	0.00000	&	1.24	\\
SDSS J1320+1644	&	200.2465628	&	16.7340505	&	0.03072	&	0.96907	&	0.00000	&	0.00003	&	0.00000	&	8.59	\\
SDSS J1322+1052	&	200.6517436	&	10.8776181	&	0.16367	&	0.83628	&	0.00000	&	0.00000	&	0.00000	&	2.00	\\
SDSS J1330+1810	&	202.5776972	&	18.1755935	&	0.76688	&	0.23279	&	0.00000	&	0.00001	&	0.00000	&	1.76	\\
SDSS1332+0347	&	203.0943281	&	3.7944053	&	0.00938	&	0.00905	&	0.00000	&	0.00009	&	0.00004	&	1.14	\\
SDSS J1334+3315	&	203.5058238	&	33.2595348	&	0.02000	&	0.79386	&	0.00000	&	0.01041	&	0.00720	&	0.83	\\
LBQS1333+0113	&	203.8949735	&	1.3015460	&	0.42463	&	0.57506	&	0.00000	&	0.00000	&	0.00000	&	1.63	\\
SDSSJ13372+6012A	&	204.3047305	&	60.2018269	&	0.00503	&	0.99495	&	0.00000	&	0.00001	&	0.00000	&	3.10	\\
SDSS J1339+1310	&	204.7797429	&	13.1776846	&	0.01348	&	0.98534	&	0.00000	&	0.00032	&	0.00000	&	1.69	\\
SDSSJ13494+1227A	&	207.3743497	&	12.4519370	&	0.45317	&	0.54683	&	0.00000	&	0.00000	&	0.00000	&	3.00	\\
SDSS1353+1138	&	208.2764435	&	11.6346476	&	0.81294	&	0.18704	&	0.00000	&	0.00000	&	0.00000	&	1.41	\\
SDSS J1400+3134	&	210.0532059	&	31.5817065	&	0.00149	&	0.99490	&	0.00000	&	0.00013	&	0.00000	&	1.74	\\
SDSSJ14002+3134	&	210.0535531	&	31.5813192	&	0.00839	&	0.98560	&	0.00000	&	0.00022	&	0.00000	&	1.74	\\
B1359+154	&	210.3982833	&	15.2233761	&	0.21200	&	0.12899	&	0.00000	&	0.11572	&	0.04951	&	1.71	\\
SDSS1402+6321	&	210.6175989	&	63.3592669	&	0.03197	&	0.00000	&	0.00180	&	0.00006	&	0.02126	&	1.35	\\
SDSSJ14050+4447A	&	211.2580744	&	44.7999512	&	0.32177	&	0.67821	&	0.00000	&	0.00002	&	0.00000	&	7.40	\\
SDSS J1405+0959	&	211.3142594	&	9.9920306	&	0.07177	&	0.92672	&	0.00000	&	0.00001	&	0.00000	&	1.98	\\
SDSS1406+6126	&	211.6034810	&	61.4447165	&	0.00423	&	0.97378	&	0.00000	&	0.01210	&	0.00000	&	1.98	\\
SDSSJ14098+3919A	&	212.4739213	&	39.3333624	&	0.00000	&	0.99201	&	0.00000	&	0.00764	&	0.00006	&	6.80	\\
HST14113+5211	&	212.8320325	&	52.1916252	&	0.00000	&	0.00000	&	0.00032	&	0.64556	&	0.30989	&	1.80	\\
J141546.24+112943.4	&	213.9426675	&	11.4953999	&	0.96943	&	0.03057	&	0.00000	&	0.00000	&	0.00000	&	1.35	\\
HST14176+5226	&	214.3989070	&	52.4462055	&	0.00000	&	0.00000	&	0.64088	&	0.00082	&	0.11522	&	2.83	\\
SDSSJ14189+2441A	&	214.7308964	&	24.6858055	&	0.01147	&	0.98760	&	0.00000	&	0.00001	&	0.00000	&	4.50	\\
B1422+231	&	216.1587845	&	22.9335307	&	1.00000	&	0.00000	&	0.00000	&	0.00000	&	0.00000	&	1.68	\\
SDSSJ14260+0719A	&	216.5177783	&	7.3238325	&	0.02721	&	0.96679	&	0.00000	&	0.00597	&	0.00000	&	4.30	\\
SDSS J1455+1447	&	223.7580380	&	14.7929914	&	0.20414	&	0.79581	&	0.00000	&	0.00000	&	0.00000	&	1.73	\\
SDSSJ15087+3328A	&	227.1758283	&	33.4673933	&	0.35610	&	0.64390	&	0.00000	&	0.00000	&	0.00000	&	2.90	\\
SDSS J1515+1511	&	228.9108052	&	15.1933151	&	0.18959	&	0.80792	&	0.00000	&	0.00003	&	0.00000	&	2.00	\\
SBS1520+530	&	230.4368204	&	52.9134682	&	0.58626	&	0.41373	&	0.00000	&	0.00001	&	0.00000	&	1.59	\\
SDSSJ15247+4409	&	231.1900916	&	44.1637501	&	0.28051	&	0.00485	&	0.00001	&	0.00007	&	0.00081	&	1.70	\\
SDSS J1527+0141	&	231.8338784	&	1.6943353	&	0.11274	&	0.88725	&	0.00000	&	0.00000	&	0.00000	&	2.58	\\
SDSS J1529+1038	&	232.4120988	&	10.6344195	&	0.29694	&	0.70278	&	0.00000	&	0.00000	&	0.00000	&	1.27	\\
SDSSJ15306+5304A	&	232.6606914	&	53.0677617	&	0.00052	&	0.99830	&	0.00000	&	0.00035	&	0.00000	&	4.10	\\
HST15433+5352	&	235.8370915	&	53.8645264	&	0.00000	&	0.01033	&	0.00099	&	0.01248	&	0.01926	&	1.18	\\
MG1549+3047	&	237.3013867	&	30.7879099	&	0.00013	&	0.00000	&	0.00002	&	0.00000	&	0.00043	&	1.70	\\
SDSSJ16002+0000	&	240.0645942	&	0.0126311	&	0.22794	&	0.77205	&	0.00000	&	0.00001	&	0.00000	&	1.80	\\
B1600+434	&	240.4187611	&	43.2796578	&	0.00189	&	0.00220	&	0.00011	&	0.00082	&	0.00486	&	1.40	\\
SDSSJ16060+2900A	&	241.5117058	&	29.0135566	&	0.24344	&	0.75656	&	0.00000	&	0.00000	&	0.00000	&	3.40	\\
SDSS J1620+1203	&	245.1089177	&	12.0616814	&	0.26150	&	0.68051	&	0.00000	&	0.00251	&	0.00000	&	2.77	\\
1WGAJ16290+3724A	&	247.2608250	&	37.4085587	&	0.16648	&	0.83351	&	0.00000	&	0.00000	&	0.00000	&	4.30	\\
PMNJ1632-0033	&	248.2403586	&	-0.5558697	&	0.00504	&	0.31223	&	0.00000	&	0.05356	&	0.00759	&	1.47	\\
FBQ1633+3134	&	248.4541069	&	31.5699816	&	0.89503	&	0.10497	&	0.00000	&	0.00000	&	0.00000	&	0.75	\\
SDSSJ16351+2911A	&	248.7922560	&	29.1890689	&	0.01140	&	0.98689	&	0.00000	&	0.00027	&	0.00000	&	4.90	\\
KP1634.9+26.7A	&	249.2538546	&	26.6027532	&	0.00003	&	0.99978	&	0.00000	&	0.00017	&	0.00000	&	3.80	\\
QJ1643+3156B	&	250.7974541	&	31.9390721	&	0.12417	&	0.87517	&	0.00000	&	0.00000	&	0.00000	&	2.30	\\
SDSS1650+4251	&	252.6810110	&	42.8637037	&	0.26140	&	0.73860	&	0.00000	&	0.00000	&	0.00000	&	1.23	\\
MG1654+1346	&	253.6741318	&	13.7725911	&	0.00030	&	0.00000	&	0.08538	&	0.00089	&	0.25019	&	2.10	\\
SDSSJ17232+5904A	&	260.8225806	&	59.0795939	&	0.05343	&	0.94657	&	0.00000	&	0.00000	&	0.00000	&	3.70	\\
B2108+213	&	317.7255259	&	21.5161064	&	0.89651	&	0.00000	&	0.00647	&	0.00015	&	0.00446	&	4.57	\\
B2114+022	&	319.2115894	&	2.4295637	&	0.56744	&	0.07080	&	0.00000	&	0.00001	&	0.00001	&	1.31	\\
SDSSJ22144+1326A	&	333.6126343	&	13.4491709	&	0.00023	&	0.99124	&	0.00000	&	0.00348	&	0.00000	&	5.80	\\
Q2237+030	&	340.1259423	&	3.3584156	&	0.55174	&	0.11150	&	0.00000	&	0.00001	&	0.00000	&	1.78	\\
B2319+052	&	350.4199408	&	5.4599435	&	0.00000	&	0.00000	&	0.01405	&	0.03596	&	0.83984	&	1.36	\\
PSS2322+1944	&	350.5298481	&	19.7397163	&	0.02608	&	0.04620	&	0.01689	&	0.04352	&	0.45616	&	1.49	\\
SDSSpJ23365-0107	&	354.1489629	&	-1.1260454	&	0.36245	&	0.63754	&	0.00000	&	0.00001	&	0.00000	&	1.70	\\
SDSS J2343-0050	&	355.7997514	&	-0.8428717	&	0.34923	&	0.64959	&	0.00000	&	0.00093	&	0.00001	&	1.51	\\
Q2345+007A	&	357.0816039	&	0.9559653	&	0.00123	&	0.99781	&	0.00000	&	0.00093	&	0.00000	&	7.10	\\
\hline																								
\end{tabular}}
\caption{List of all known lenses in SDSS, along with selected membership probabilities. (\textit{Continued from previous page})}
\label{tab:knownlensscores}
\end{table*}

The discrepancy in the scores is likely due to the fact that we trained our model on blended, small separation lens systems and so the model is designed to pick out similar objects. Since the colours of deblended objects come only from a single image, they appear as stand-alone, point-like QSO, as in the bottom three images in Figure \ref{fig:knowncutouts} (a). Blended objects, such as those in the top three images of Figure \ref{fig:knowncutouts} (a), will be extended and will have the colours of quasars. Thus, our model will identify them either as lenses or extended QSO.

In some cases, lenses that are blended in SDSS will become deblended in surveys such as DES, where the seeing and image quality are better. This may pose a problem with our model as those systems will likely receive low lens membership probabilities when examined in DES. In order to capture as many lenses as possible, it becomes increasingly important to extend our training sets to include objects of all image configurations and separations.

\section{Lens Candidates}
\label{sect:candidates}

The membership probabilities produced by our model allow for numerous
methods of selecting lens candidates. The simplest method is to take
as targets the objects with the highest combined lens probabilities,
while alternative choices may place upper limits on, say, the
blue-cloud galaxy probabilities. After a list of targets is compiled,
candidate selection can be based for example on visual inspection of
the images, machine learning pixel based techniques \citep{Agn++15a},
or fast lens modeling \citep{Marshall:2009p593,Cha++15}.

\begin{figure}
\centering
\includegraphics[width=3in]{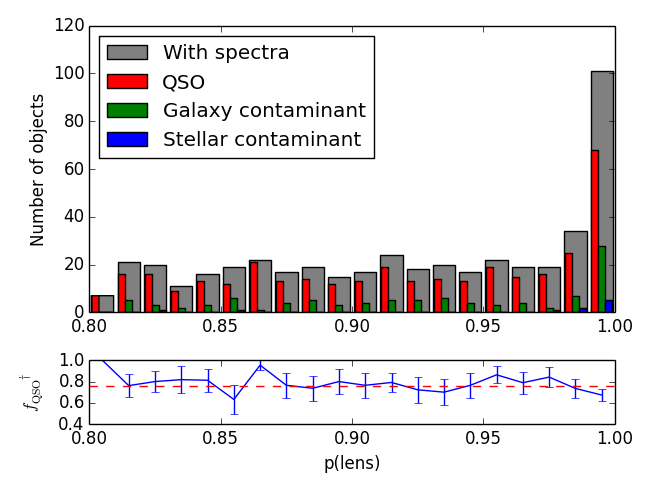}
\caption{Distribution of the 458 objects with spectra from our list of 2000 objects receiving the highest $p$(lens) scores. The bottom panel shows the fraction of objects with spectra that are, indeed, QSOs. The dashed red line indicates the mean fraction.\hspace{\textwidth}
$^\dag$ We define $f_{\text{QSO}} = N_\text{objects with QSO specra}/N_\text{objects with specra}$}
\label{fig:spectrabreakdown}
\end{figure}

As an illustration, we make a simple list of candidates by first examining the objects assigned the highest combined lens probabilities, summed over all three implementations. Taking the top 2000 candidates, we select those with available SDSS spectroscopy, in order to simulate a potential follow-up campaign. Of the 2000 objects in the list, 458 have spectra. We identify those with `QSO' or `Galaxy AGN' spectra as QSOs, those with `Galaxy,' `Galaxy starburst,' or `Galaxy starforming' spectra as Galaxy contaminants, and those with stellar spectra as stellar contaminants. The distribution of all objects with spectra is shown in the top frame of Figure \ref{fig:spectrabreakdown}. As hoped, nearly 80\% of the selected objects with spectra are quasars, with the majority of contaminants accounted for by other galaxies. 

\newcommand{\scalepiccand}{0.7in}
\begin{figure*}
\centering
\includegraphics[width=\scalepiccand]{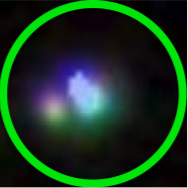}
\includegraphics[width=\scalepiccand]{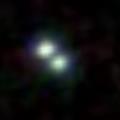}
\includegraphics[width=\scalepiccand]{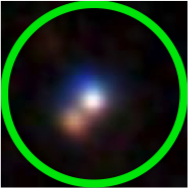}
\includegraphics[width=\scalepiccand]{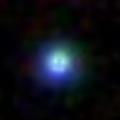}
\includegraphics[width=\scalepiccand]{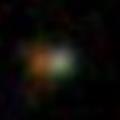}
\includegraphics[width=\scalepiccand]{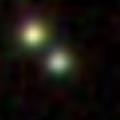}
\includegraphics[width=\scalepiccand]{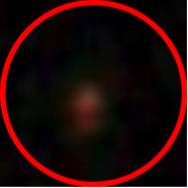}
\includegraphics[width=\scalepiccand]{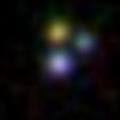}
\includegraphics[width=\scalepiccand]{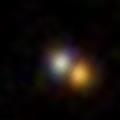}
\includegraphics[width=\scalepiccand]{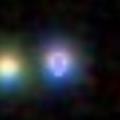}
\includegraphics[width=\scalepiccand]{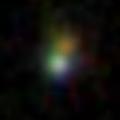}
\includegraphics[width=\scalepiccand]{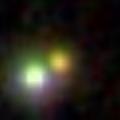}
\includegraphics[width=\scalepiccand]{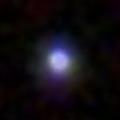}
\includegraphics[width=\scalepiccand]{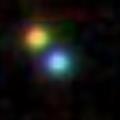}
\includegraphics[width=\scalepiccand]{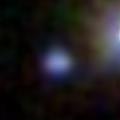}
\includegraphics[width=\scalepiccand]{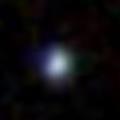}
\includegraphics[width=\scalepiccand]{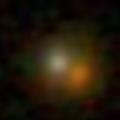}
\includegraphics[width=\scalepiccand]{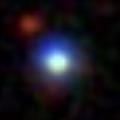}
\includegraphics[width=\scalepiccand]{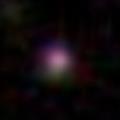}
\includegraphics[width=\scalepiccand]{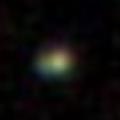}
\includegraphics[width=\scalepiccand]{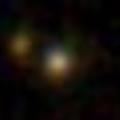}
\includegraphics[width=\scalepiccand]{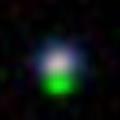}
\includegraphics[width=\scalepiccand]{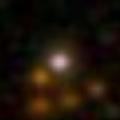}
\includegraphics[width=\scalepiccand]{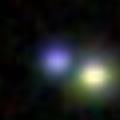}
\includegraphics[width=\scalepiccand]{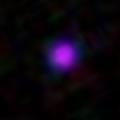}
\includegraphics[width=\scalepiccand]{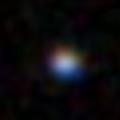}
\includegraphics[width=\scalepiccand]{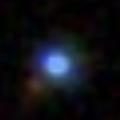}
\includegraphics[width=\scalepiccand]{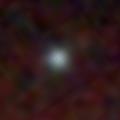}
\includegraphics[width=\scalepiccand]{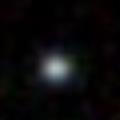}
\includegraphics[width=\scalepiccand]{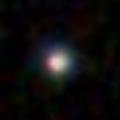}
\includegraphics[width=\scalepiccand]{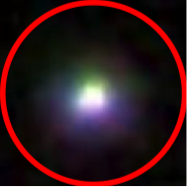}
\includegraphics[width=\scalepiccand]{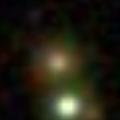}
\includegraphics[width=\scalepiccand]{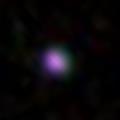}
\includegraphics[width=\scalepiccand]{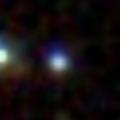}
\includegraphics[width=\scalepiccand]{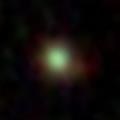}
\includegraphics[width=\scalepiccand]{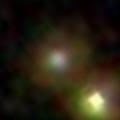}
\includegraphics[width=\scalepiccand]{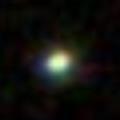}
\includegraphics[width=\scalepiccand]{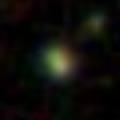}
\includegraphics[width=\scalepiccand]{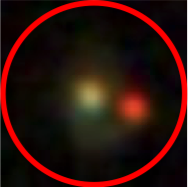}
\includegraphics[width=\scalepiccand]{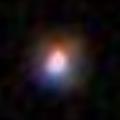}
\includegraphics[width=\scalepiccand]{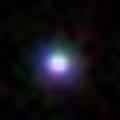}
\includegraphics[width=\scalepiccand]{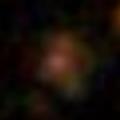}
\includegraphics[width=\scalepiccand]{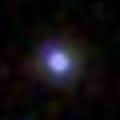}
\caption{Image cutouts of the 43 SDSS objects receiving the highest combined lensed quasar probability, corresponding to those listed in Table \ref{tab:candcoord}. All the objects are spectroscopically confirmed to host an active galactic nucleus. The first object in Table \ref{tab:candcoord} corresponds to the top left image. Subsequent objects are in order of left to right first and then top to bottom. Each image is 12 arcseconds on a side with North up and East to the left. The five encircled objects already have HST imaging counterparts (two known lenses in green and three singly imaged AGN in red), as discussed in the text and fig.~\ref{fig:HSTimages}.}
\label{fig:candidatecutouts}
\end{figure*}

\begin{table*}
\centering
\setlength{\tabcolsep}{4.5pt}
\begin{tabular}{r r r c c c c c c c c c c c c c}
\hline
ra	&	dec	&	u	&	g	&	r	&	i	&	z	&	W1	&	W2	&	psf\_g	&	psf\_r	&	psf\_i	&	W3	&	redshift	&	$p$(lens)	\\
\hline																													
26.3191345	&	-9.7547843	&	17.79	&	17.28	&	17.07	&	16.63	&	16.85	&	13.47	&	12.63	&	17.05	&	16.89	&	16.66	&	9.52	&	2.725	&	0.92514	\\
322.2760283	&	-7.2703745	&	19.25	&	19.01	&	18.67	&	19.45	&	18.74	&	14.97	&	13.68	&	19.03	&	18.65	&	19.04	&	11.30	&	1.215	&	0.90465	\\
203.8949676	&	1.3015746	&	18.74	&	18.81	&	18.70	&	17.97	&	18.20	&	14.06	&	12.79	&	18.57	&	18.38	&	17.79	&	9.61	&	1.576	&	0.99458	\\
39.0260148	&	-8.7158960	&	17.59	&	17.38	&	17.32	&	17.36	&	17.32	&	13.52	&	12.18	&	17.35	&	17.22	&	17.33	&	9.29	&	0.893	&	0.82064	\\
356.5790860	&	-9.5192272	&	20.39	&	19.80	&	18.98	&	18.91	&	19.07	&	14.26	&	13.39	&	19.96	&	19.15	&	19.17	&	10.94	&	1.111	&	0.99347	\\
245.3698819	&	-0.6473633	&	20.00	&	19.58	&	18.88	&	19.16	&	18.63	&	14.66	&	13.71	&	19.61	&	18.91	&	19.17	&	11.11	&	1.061	&	0.97723	\\
24.8632464	&	1.2630244	&	22.06	&	21.65	&	20.58	&	19.69	&	19.84	&	15.13	&	13.81	&	22.35	&	21.06	&	20.27	&	9.39	&	0.230	&	0.96277	\\
232.9027584	&	4.2669427	&	18.92	&	18.25	&	18.68	&	18.43	&	18.17	&	14.98	&	14.03	&	18.81	&	19.23	&	19.00	&	11.01	&	2.000	&	0.95502	\\
257.0625197	&	34.8380944	&	19.37	&	19.15	&	19.69	&	19.12	&	18.92	&	14.47	&	13.17	&	18.95	&	18.93	&	18.53	&	10.29	&	1.600	&	0.90956	\\
189.0097472	&	-3.5249859	&	17.09	&	16.96	&	16.92	&	16.68	&	16.74	&	13.58	&	12.28	&	16.98	&	16.91	&	16.66	&	9.14	&	1.824	&	0.90437	\\
\\																													
36.7783671	&	-9.1214118	&	19.79	&	19.16	&	18.60	&	18.59	&	18.34	&	13.87	&	12.63	&	19.48	&	18.97	&	19.04	&	9.84	&	0.960	&	0.82243	\\
191.2738981	&	3.1463089	&	20.57	&	19.55	&	18.36	&	17.95	&	17.71	&	12.91	&	11.78	&	19.62	&	18.43	&	18.02	&	9.10	&	1.097	&	0.81733	\\
115.3586716	&	42.2585368	&	17.62	&	17.49	&	17.53	&	17.13	&	17.05	&	13.96	&	12.59	&	17.53	&	17.60	&	17.12	&	9.27	&	1.850	&	0.98504	\\
18.6724107	&	-9.2997739	&	18.78	&	18.38	&	18.30	&	18.43	&	19.28	&	14.11	&	13.22	&	18.36	&	18.31	&	18.44	&	10.91	&	0.763	&	0.98000	\\
211.1369562	&	1.1384257	&	18.84	&	18.56	&	18.77	&	18.77	&	19.45	&	14.62	&	13.56	&	18.57	&	18.70	&	18.73	&	10.86	&	0.634	&	0.92927	\\
26.8469730	&	14.7225253	&	18.75	&	18.53	&	18.03	&	18.11	&	17.93	&	13.62	&	12.58	&	18.62	&	18.34	&	18.28	&	9.73	&	0.433	&	0.91726	\\
26.0004127	&	13.2088043	&	19.46	&	19.07	&	18.48	&	18.21	&	17.74	&	13.95	&	13.34	&	19.61	&	19.16	&	18.93	&	10.82	&	0.289	&	0.87390	\\
158.1535243	&	-1.1030481	&	17.09	&	16.99	&	16.77	&	16.72	&	16.80	&	13.43	&	12.00	&	17.05	&	16.77	&	16.71	&	8.97	&	1.263	&	0.86483	\\
33.2371116	&	-9.4852927	&	19.72	&	19.47	&	19.43	&	18.80	&	18.55	&	14.43	&	13.47	&	19.65	&	19.65	&	19.06	&	10.66	&	0.415	&	0.84925	\\
211.5935059	&	-1.2085686	&	20.43	&	19.58	&	18.96	&	18.72	&	18.68	&	14.44	&	12.87	&	19.74	&	19.12	&	18.88	&	9.61	&	1.154	&	0.83553	\\
\\																													
57.1105520	&	-0.8798747	&	20.84	&	20.41	&	19.41	&	19.07	&	18.53	&	14.88	&	14.12	&	20.67	&	19.76	&	19.43	&	12.10	&	0.266	&	0.81980	\\
352.8876504	&	-9.0462571	&	18.70	&	18.00	&	18.06	&	17.78	&	17.31	&	14.37	&	13.39	&	18.05	&	18.20	&	17.90	&	9.80	&	2.455	&	0.93794	\\
10.4610855	&	0.1244942	&	20.64	&	19.97	&	20.24	&	18.78	&	18.56	&	13.52	&	12.57	&	19.96	&	20.22	&	18.81	&	9.75	&	0.456	&	1.00000	\\
200.7151939	&	0.7818887	&	18.65	&	18.42	&	18.76	&	18.53	&	19.38	&	13.64	&	12.98	&	18.42	&	18.65	&	18.49	&	10.75	&	0.520	&	0.99882	\\
113.1172315	&	38.4402326	&	19.06	&	18.85	&	20.84	&	18.57	&	18.43	&	14.83	&	13.38	&	19.09	&	21.38	&	18.87	&	10.72	&	1.138	&	0.99872	\\
146.0592382	&	1.0485406	&	19.76	&	20.13	&	18.73	&	18.53	&	20.21	&	13.98	&	12.95	&	19.84	&	18.70	&	18.55	&	10.24	&	0.693	&	0.98932	\\
6.1838017	&	0.5393128	&	17.18	&	16.93	&	16.58	&	16.89	&	16.40	&	12.67	&	11.57	&	16.91	&	16.77	&	16.90	&	9.28	&	0.402	&	0.98864	\\
123.6649547	&	47.3068383	&	20.09	&	20.85	&	20.18	&	20.15	&	20.63	&	14.52	&	13.60	&	20.23	&	19.86	&	19.84	&	10.64	&	0.782	&	0.98531	\\
202.8579850	&	0.7374107	&	19.01	&	18.71	&	19.72	&	18.20	&	17.99	&	16.10	&	14.45	&	18.85	&	19.98	&	18.39	&	10.95	&	2.018	&	0.98444	\\
157.5410952	&	1.4848465	&	18.99	&	18.74	&	18.32	&	17.89	&	18.21	&	14.49	&	13.14	&	18.77	&	18.32	&	17.79	&	10.58	&	1.277	&	0.95529	\\
\\																													
12.6721473	&	-9.4847679	&	16.63	&	16.24	&	15.81	&	15.57	&	15.33	&	13.38	&	12.39	&	16.31	&	15.79	&	15.52	&	9.78	&	1.192	&	0.94780	\\
163.5762432	&	0.1043169	&	19.52	&	19.26	&	18.71	&	18.33	&	17.81	&	13.92	&	13.32	&	19.75	&	19.30	&	18.97	&	10.79	&	0.349	&	0.93549	\\
124.4711822	&	45.8888824	&	19.31	&	19.19	&	19.44	&	18.77	&	18.77	&	15.69	&	14.31	&	19.28	&	19.73	&	18.91	&	11.21	&	1.742	&	0.90558	\\
43.9399370	&	-0.8650154	&	19.62	&	19.38	&	19.25	&	19.98	&	19.70	&	14.68	&	13.79	&	19.41	&	19.27	&	19.68	&	11.31	&	0.751	&	0.89951	\\
112.6878885	&	36.5752726	&	19.93	&	19.06	&	18.37	&	18.34	&	18.02	&	13.77	&	12.46	&	19.06	&	18.36	&	18.39	&	9.78	&	1.063	&	0.89873	\\
122.4955314	&	45.7172141	&	19.39	&	19.14	&	18.68	&	18.39	&	17.83	&	14.30	&	13.60	&	19.75	&	19.51	&	19.30	&	10.87	&	0.366	&	0.89737	\\
358.1587047	&	1.0978856	&	18.98	&	18.16	&	17.54	&	17.23	&	17.07	&	14.05	&	12.84	&	18.11	&	17.54	&	17.21	&	9.25	&	2.994	&	0.88778	\\
209.9342104	&	1.4693924	&	19.84	&	19.67	&	18.97	&	18.85	&	18.74	&	14.75	&	13.50	&	19.88	&	19.30	&	19.27	&	10.06	&	1.096	&	0.88693	\\
39.2512724	&	-1.0251575	&	19.86	&	19.35	&	18.68	&	18.47	&	18.00	&	14.39	&	13.80	&	19.71	&	19.06	&	18.86	&	10.88	&	0.344	&	0.88225	\\
15.0132242	&	15.8495361	&	17.66	&	17.32	&	17.07	&	16.59	&	16.70	&	13.68	&	12.73	&	17.39	&	17.15	&	16.65	&	9.32	&	0.109	&	0.86729	\\
\\																													
14.6031392	&	0.6870584	&	17.02	&	17.05	&	16.94	&	16.65	&	16.53	&	13.78	&	12.53	&	17.09	&	17.00	&	16.69	&	9.29	&	1.921	&	0.85083	\\
197.0064644	&	0.0958673	&	20.92	&	20.12	&	19.44	&	18.69	&	18.54	&	14.43	&	13.53	&	21.07	&	20.64	&	19.92	&	10.42	&	0.480	&	0.83156	\\
33.2483206	&	-0.5081826	&	17.90	&	17.65	&	17.67	&	17.49	&	17.08	&	14.34	&	13.21	&	17.70	&	17.73	&	17.56	&	9.94	&	0.395	&	0.83109	\\
	\hline
\end{tabular}
\caption{Coordinates and colours of the 43 SDSS objects selected by visual inspection. The candidates are ranked primarily by visual inspection score and secondarily by lens score.}
\label{tab:candcoord}
\end{table*}

\begin{figure}
\centering
\includegraphics[width=3in]{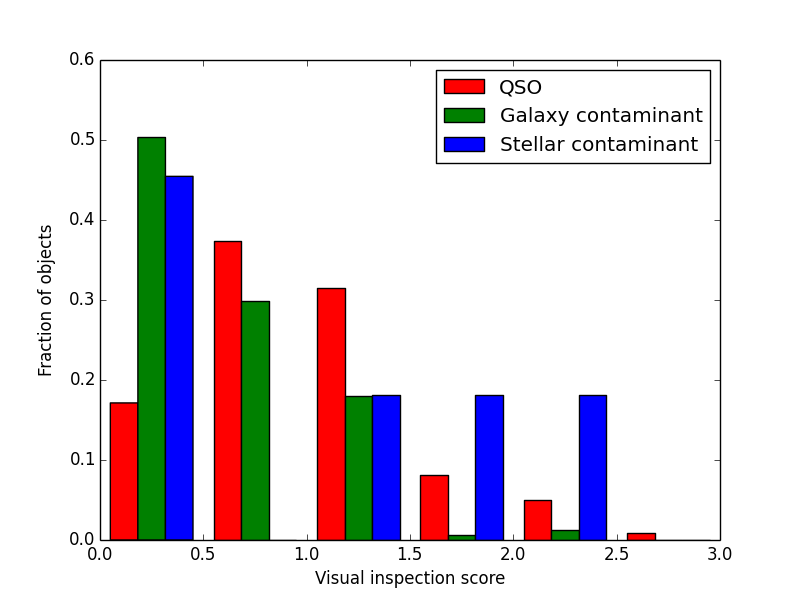}
\caption{Distributions of visual inspection scores, split into QSOs, galaxy contaminants, and stellar contaminants. Note that the stellar contaminants come from a small sample size of $N=11$.}
\label{fig:ourrankings}
\end{figure}

Taking the list of objects with spectra, we visually inspect them and assign a score of 0-3, where 0 corresponds to `not a lens' and 3 corresponds to `likely a lens.' We assign the rankings blind to the spectra so as to not be biased in our score assignments. We use the spectra afterwards as a check on our visual inspection step to ensure that we do not assign high scores to the contaminant classes. The distribution of our scores is shown in Figure \ref{fig:ourrankings}. We first note the difficulty in distinguishing between QSOs and stellar contaminants. However, this is a very small sample of our list, making up only $\sim$2\% of all objects. We are much better at identifying galaxy contaminants, assigning scores less than 3\% of them scores greater than 1.5.

From our visual inspection scores, we have 43 objects with score greater than 1.5, after eliminating all contaminants based on the spectroscopic information. These are listed in Table \ref{tab:candcoord}, sorted first by visual inspection score, and secondarily by the lens probability assigned by our model. Image cutouts for of the objects are shown in Figure \ref{fig:candidatecutouts}.

Of the 43 selected candidates, 5 objects have imaging available in the Hubble Legacy Archive \citep{hst1,hst2,hst3,hst4,hst5}. The images of these objects are shown below their SDSS counterparts in \ref{fig:HSTimages}. Of the five objects, two are known lensed quasars while the others are singly imaged AGN. The two known lenses correspond to rank 1 and 3 in our list of candidates, providing more confidence in our visual inspection step.

\newcommand{\scalepichst}{1.3in}
\begin{figure*}
\includegraphics[width=\scalepichst]{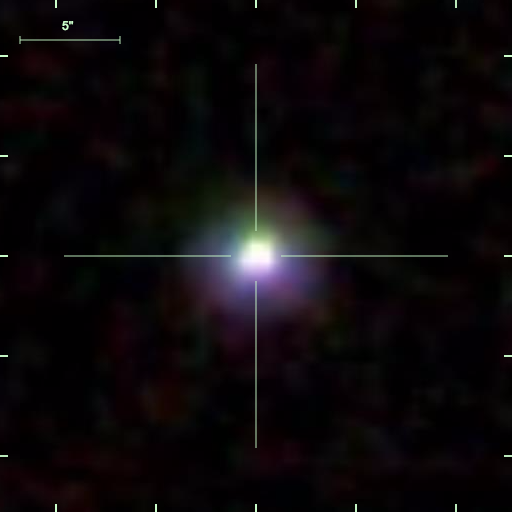}	
\includegraphics[width=\scalepichst]{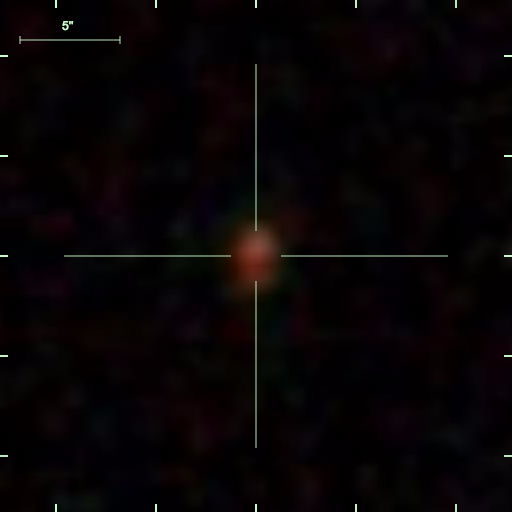}	
\includegraphics[width=\scalepichst]{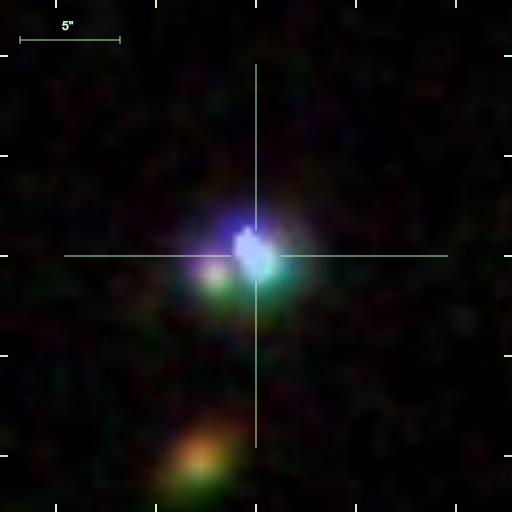}	
\includegraphics[width=\scalepichst]{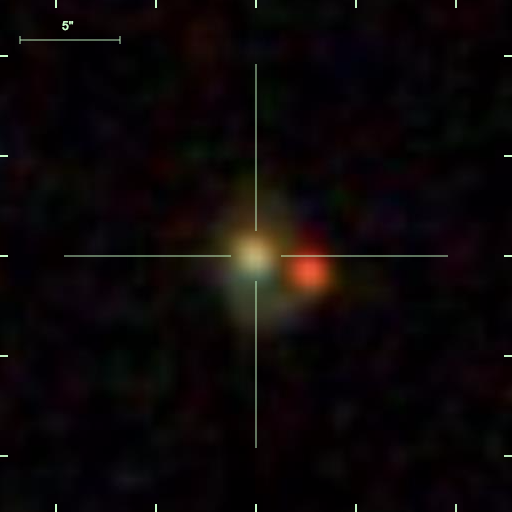}	
\includegraphics[width=\scalepichst]{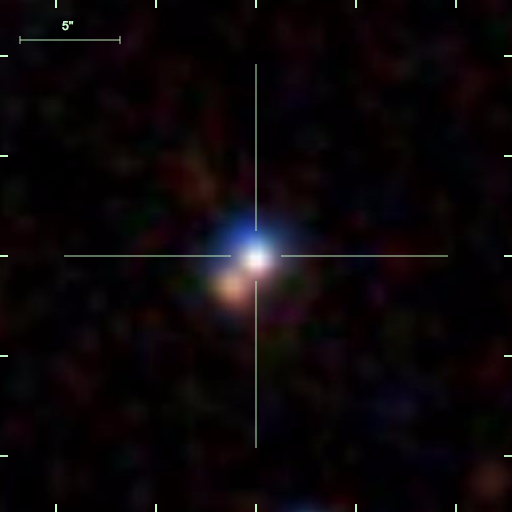}	
\includegraphics[width=\scalepichst]{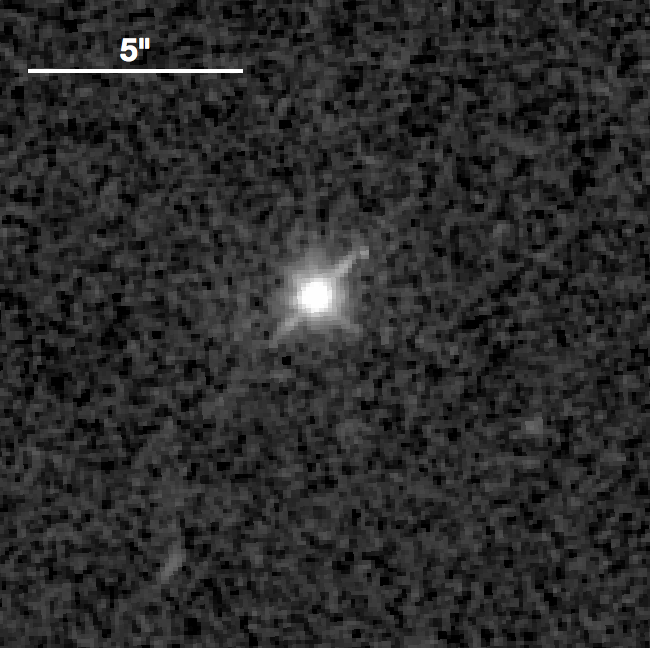}	
\includegraphics[width=\scalepichst]{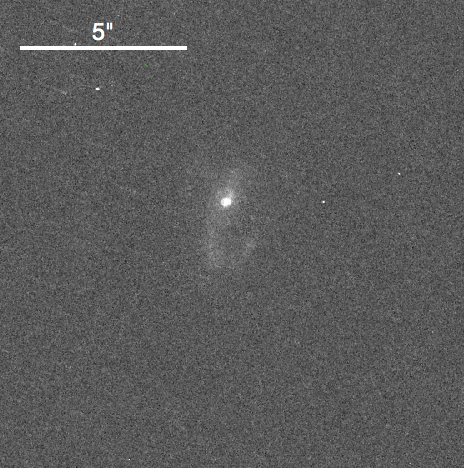}
\includegraphics[width=\scalepichst]{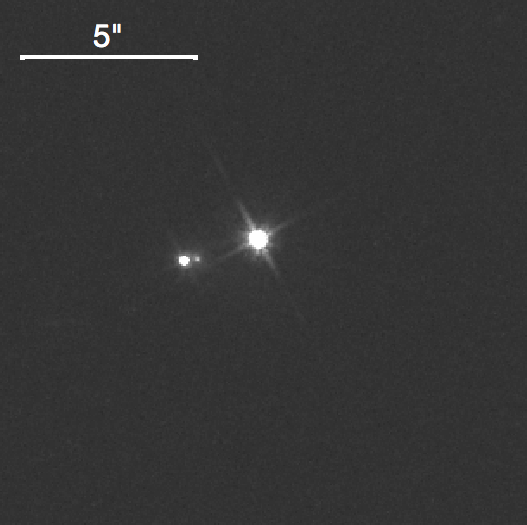}
\includegraphics[width=\scalepichst]{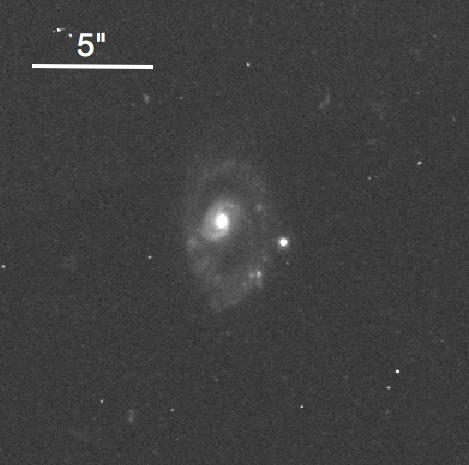}
\includegraphics[width=\scalepichst]{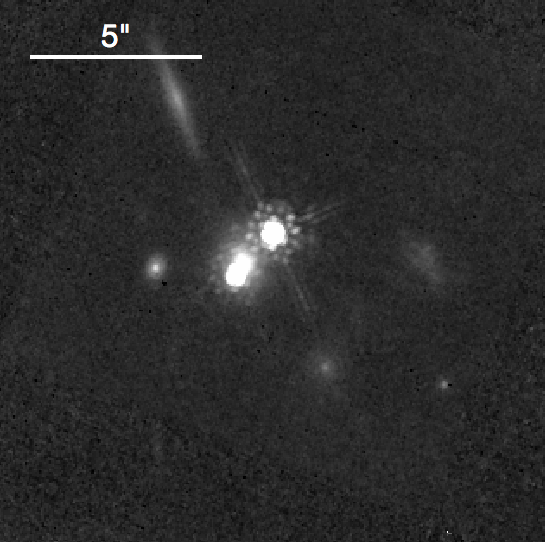}
\caption{5 of our top 43 candidates with imaging data available in the Hubble Legacy Archive. From left to right, the objects correspond to 31st, 7th, 1st, 39, and 3rd objects in Table \ref{tab:candcoord}. The middle and rightmost objects are both lenses.}
\label{fig:HSTimages}
\end{figure*}

It is important to remember that we only used the probabilities from about 10\% of the objects in SDSS passing our initial colour cuts (Section \ref{sect:largesdss}), so we expected to recover at best 10\% of the known lenses in the SDSS footprint. In our list of 2000 objects with the highest lens probability, the lowest probability was near 0.8. In a full examination of the entire survey, a combination of cuts would be needed: more demanding cuts at query level, e.g. on the limiting $W2$ magnitude and WISE colours; upper cuts in membership probabilities relative to other classes; a cutoff in the minimum $p(\rm{lens})$ score; and possibly a re-modulation of other membership probabilities, such as extended quasars at $1.2<z<1.75,$ to account for the artificial clustering around specific classes.

\section{Summary and Conclusions}
\label{sect:concl}

We have demonstrated the use of Gaussian mixture models as a possible method of searching for strongly lensed quasars using purely photometric data. We begin our search by first identifying different types of possible contaminants. For each of these classes, we develop training sets either with simulated objects in the case of rare objects such as lenses, or with real objects from SDSS for the more populated classes. Using these classes with our model along with the Expectation Maximization algorithm, we fit our model to the data in order to best describe the different populations. Our main results can be summarized as follows:

\begin{enumerate}
\item Our Gaussian mixture model is capable of discriminating between our chosen classes, and the performance improves by adding features. After training, the model is flexible enough to adjust itself to real and larger SDSS datasets, and further improve its performance.

\item When translated to VST-ATLAS, the model trained on SDSS is still able to sort most objects into their proper classes. 
However, we find that the \texttt{psf} and \texttt{model} magnitudes, which prove useful in the SDSS implementation, cannot be easily replaced by the \texttt{AperMag3} and \texttt{AperMag6} magnitudes in VST-ATLAS.

\item When tested on the known lenses in SDSS, the model typically assigns either very high or very low lens probabilities. The model performs well on small separation blended systems that are the main focus of our search. Conversely, lenses that receive low probabilities typically receive high point-like QSO probabilities and are preferentially lenses with large image separation. In future implementations, one could include additional training sets designed to find this additional class of lenses. 

\item Of the objects receiving high lens probabilities, roughly 80\% are QSOs and 20\% are either galaxy or stellar contaminants. An additional classification step is necessary before they can be considered viable lens candidates in order to minimize contamination. We illustrate this step using visual classification, and produce a list of 43 lensed quasar candidates selected from 10\% of SDSS, all of which are known to contain an active galactic nucleus from SDSS spectroscopy. Five of the candidates happen to have archival HST images, including two known lenses.  

\end{enumerate}

A word of caution is also in order. The expectation maximization algorithm used on a Gaussian mixture model remains a semi-supervised machine learning method. The Gaussian parameters adjust themselves based on the data, but the method requires user supervision to ensure that the classes assigned by the model do, indeed, correspond to the classes they were programmed to describe. Since the EM algorithm can only maximize the likelihood function and is not penalized for misclassifying objects in the training set, it is possible that one of the Gaussians might describe a different class of objects than it was initially set to describe.

In conclusion, our study demonstrates the power of Gaussian Mixture
Model for selecting samples of lens quasar candidates with high
purity, suitable for spectroscopic and/or high resolution imaging
follow-up. The strategy illustrated here, however, is by no means
unique and different choices and improvements are possible. For
example, one could include stricter colour cuts before the visual
inspection steps in an effort to increase the purity of the sample. As
we showed with the known lenses in SDSS, one can reduce the total
sample by an order of magnitude while losing only small fraction of
the true lenses. Furthermore, we have seen that increasing features
and the proper selection of features can have a drastic effect on the
performance of the model. By extending the model to higher dimensions,
it will be possible to improve the selection, but with increased
computational complexity. Alternatively, the visual inspection step
could be replaced with pixel based machine learning techniques
\citep[e.g.][]{Agn++15a} or model based techniques \cite{Marshall:2009p593,Cha++15}.

\section*{Acknowledgments}
We are indebted to Philip J. Marshall for the suggestion of using
population-based approaches and for insightful discussions during the
development of the method that helped improve this paper. We thank our
colleagues in the STrong lensing Insights into the Dark Energy Survey
(STRIDES\footnote{STRIDES is a Dark Energy Survey Broad External
Collaboration; PI: Treu. \url{http://strides.astro.ucla.edu}})
collaboration for many stimulating conversations on lens finding. We
acknowledges support by the Packard Foundations through a Packard
Research Fellowship and by the National Science Foundation through
grant AST-1450141. 
The code used to implement gaussian mixture models as well as our tables of training sets are available through the github repository \url{https://github.com/tommasotreu/WAT_GMM}. Please email TT to request access.

\bibliographystyle{mnras}
\bibliography{references} 

\appendix
\onecolumn
\section{Expectation Maximization algorithm}
\label{sect:EMap}

Suppose our data is made up of $N$ objects, each with $P$ features. We
will call this $\bm{x} = x_{ip}, i \in [1, N], p \in [1, P]$. We wish
to describe our data as a superposition of $K$ populations, each
described by a different parent distribution function (PDF),
$p(\bm{x}_i | \bm{\theta}_k)$, where $\bm{\theta}_k$ are the
parameters defining the PDF. We use
Gaussians as our PDFs, so
\begin{align}
\nonumber p(\bm{x}_i | \bm{\theta}_k) &= \frac{1}{(2\pi)^{P/2} |\bm{\Sigma}_k|^{1/2}} e^{-\frac{1}{2}(\bm{x}_i-\bm{\mu}_k)^T\bm{\Sigma}_k^{-1}(\bm{x}_i-\bm{\mu}_k)} \\
&\equiv f(\bm{x}_i | \bm{\theta}_k).
\label{eq:pdf}
\end{align}
Here, $\bm{\theta} = \{\bm{\mu},\bm{\Sigma},\bm{\alpha}\}$, where $\bm{\mu} = \mu_{kp}$ are the means, $\bm{\Sigma} = \Sigma_{kp_1p_2}$ are the covariance matrices, and $\bm{\alpha}=\alpha_k$ are the Gaussian weights. 

We can construct a posterior function
\begin{equation}
p(\bm{\theta} | \{\bm{x}_i\}) \propto p(\{\bm{x}_i\} | \bm{\theta}) p(\bm{\theta})
\end{equation}
which we want to maximize by finding
\begin{align}
\bm{\theta}_\text{best} = \underset{\bm{\theta}}{\arg \max} \,\,p(\{\bm{x}_i\} | \bm{\theta}).
\end{align}
We can maximize the likelihood function
\begin{equation}
p(\{\bm{x}_i\} | \bm{\theta}) = \prod_i \sum_k \alpha_k \cdot f(\bm{x}_i | \bm{\theta}_{k})
\end{equation}
by means of the Expectation Maximization algorithm.

To begin the Expectation Maximization algorithm, we first make initial guesses for the parameters, $\bm{\theta}$. Because we know what objects we are attempting to describe, we make informed guesses based on where we expect each class to lie in feature space. In the \textit{Expectation Step}, we calculate the expected value of the likelihood function, based on the current parameters estimates. In this step, we also calculate \textit{membership probabilities}, i.e., the probability that $\bm{x}_i$ belongs to class $k$, given our current estimate of $\bm{\theta}$. These are given by 
\begin{equation}
w_{ik} = \frac{\alpha_k \cdot f(\bm{x}_i | \bm{\theta_k})}{\sum_{k'} \alpha_{k'} \cdot f(\bm{x}_i | \bm{\theta_{k'}})}
\end{equation}
 so that $\sum_k w_{ik} = 1$.
 
In the \textit{Maximization Step}, we want to adjust the parameters $\bm{\theta}$ so as to maximize the likelihood function. This is done according to the following equations:
\begin{align}
\alpha_k^{new} &= \frac{\sum_i w_{ik}}{N}\\
\bm\mu_k^{new} &= \frac{1}{\sum_i w_{ik}}\sum_i w_{ik} \cdot \bm{x}_i\\
\bm\Sigma_k^{new} &= \frac{1}{\sum_i w_{ik}}\sum_i w_{ik} \cdot (\bm{x}_i - \bm\mu_k^{new})(\bm{x}_i - \bm\mu_k^{new})^T
\end{align}
Using the newly calculated parameters, we repeat the process, iterating the expectation and maximization steps until satisfactory convergence.

\label{lastpage}
\end{document}